\documentclass[transmag,10pt]{IEEEtran}
\makeatletter

\newtheorem{lemma}{Lemma}
\IEEEoverridecommandlockouts
\usepackage{cite}
\usepackage{amsmath,amssymb,amsfonts,float,graphicx,steinmetz,bm,subfigure,multicol,multirow,url,dblfloatfix}
\usepackage[linesnumbered,ruled,vlined]{algorithm2e}
\SetKwInput{KwInput}{Input}                
\SetKwInput{KwOutput}{Output}    
\usepackage{mathtools}

\usepackage{textcomp}
\usepackage{tikz,xcolor}
\DeclareMathSizes{10}{9.5}{6}{5}

\definecolor{lime}{HTML}{A6CE39}
\DeclareRobustCommand{\orcidicon}{%
	\begin{tikzpicture}
	\draw[lime, fill=lime] (0,0) 
	circle [radius=0.16] 
	node[white] {{\fontfamily{qag}\selectfont \tiny ID}};
	\draw[white, fill=white] (-0.0625,0.095) 
	circle [radius=0.007];
	\end{tikzpicture}
	\hspace{-2mm}
}

\foreach \x in {A, ..., Z}{%
	\expandafter\xdef\csname orcid\x\endcsname{\noexpand\href{https://orcid.org/\csname orcidauthor\x\endcsname}{\noexpand\orcidicon}}
}

\def\BibTeX{{\rm B\kern-.05em{\sc i\kern-.025em b}\kern-.08em T\kern-.1667em\lower.7ex\hbox{E}\kern-.125emX}}

\begin{document}             
\title{
Channel Estimation for Large Intelligent Surface Aided MISO Communications: From LMMSE to Deep Learning Solutions
\thanks{This work has been submitted to the IEEE for possible publication. Copyright may be transferred without notice, after which this version may no longer be accessible}
 \thanks{This work was supported by the General Research Fund of the Hong Kong Research Grants Council (grant number 16202918). It was also supported by the Hong Kong PhD Fellowship Scheme (PF17-00157).}
\thanks{A preliminary version of this work was presented at the 2020 IEEE International Symposium on Personal, Indoor and Mobile Radio Communications (PIMRC) \cite{kunduLIS_pimrc}.}

}

 \author{
 Neel Kanth Kundu\orcidA{}, {\em Graduate Student Member, IEEE}, and
 Matthew R. McKay\orcidB{}, {\em Senior Member, IEEE}
 \thanks{
 Neel Kanth Kundu and Matthew R. McKay are with the Department of
 Electronic and Computer Engineering, The Hong Kong University of
 Science and Technology, Clear Water Bay, Kowloon, Hong Kong
 (e-mail: {\tt nkkundu@connect.ust.hk}, {\tt m.mckay@ust.hk}).}
 }

\IEEEtitleabstractindextext{
\begin{abstract}
We consider multi-antenna wireless systems aided by large intelligent surfaces (LIS). LIS presents a new physical layer technology for improving coverage and energy efficiency by intelligently controlling the propagation environment. In practice however, achieving the anticipated gains of LIS requires accurate channel estimation. Recent attempts to solve this problem have considered the least-squares (LS) approach, which is simple but also sub-optimal. The optimal channel estimator, based on the minimum mean-squared-error (MMSE) criterion, is challenging to obtain and is non-linear due to the non-Gaussianity of the effective channel seen at the receiver. Here we present approaches to approximate the optimal MMSE channel estimator. As a first approach, we analytically develop the best linear estimator, the LMMSE, together with a corresponding majorization-minimization based algorithm designed to optimize the LIS phase shift matrix during the training phase. This estimator is shown to yield improved accuracy over the LS approach by exploiting second-order statistical properties of the wireless channel and the noise.  To further improve performance and better approximate the globally-optimal MMSE channel estimator, we propose data-driven non-linear solutions based on deep learning.  Specifically, by posing the MMSE channel estimation problem as an image denoising problem, we propose two convolutional neural network (CNN) based methods to perform the denoising and approximate the optimal MMSE channel estimation solution.  Our numerical results show that these CNN-based estimators give superior performance compared with linear estimation approaches. They also have low computational complexity requirements, thereby motivating their potential use in future LIS-aided wireless communication systems.  
\end{abstract}

}
\maketitle
\markboth{This work has been submitted to the IEEE for possible publication. Copyright may be transferred without notice. }
\IEEEpeerreviewmaketitle

\section{Introduction}

Future communication systems will need to support billions of connected devices, while offering extremely high data rates and system reliability \cite{popovski20185g}. Meeting these challenges will require the development of innovative physical layer technologies. Various approaches are being explored, such as the development of new antenna technologies \cite{zhang2020prospective}, multiplexing schemes \cite{zhu2019millimeter}, and signaling strategies \cite{kundu2020signal,basar2020reconfigurable}. 

Large intelligent surfaces (LIS), alternatively labelled reconfigurable intelligent surfaces (RIS), are also attracting considerable attention, and are being envisioned as a promising physical layer technology for enhancing the performance of future wireless systems  \cite{risdi2019smart,di2020smart,huang2020holographic}.
LIS can potentially improve network coverage, and can enable spectral and energy efficiency gains that are commensurate with massive MIMO systems, but with much fewer antennas at the base station (BS) \cite{huang2019reconfigurable}. A distinctive feature of LIS is that they do not have active components like power amplifiers and analog-to-digital converters, but instead are composed of almost passive elements that can intelligently directionally control the propagation of impinging electromagnetic waves to improve end-to-end performance. 

Various recent contributions have considered the use of LIS to improve the performance of multiple-antenna communication systems. For such systems, improvements in network spectral efficiency and energy efficiency have been demonstrated through the use of joint active and passive beamforming at the BS and the LIS unit respectively \cite{debbahhuang2018achievable,huang2019reconfigurable, ruiwu2018intelligent,wu2019intelligent,kammoun2020asymptotic,wang2019intelligent,zhou2020spectral,kundu2020risassisted}. A common assumption of these works is that perfect channel state information is available for optimizing the beamforming vectors. In practice however, channel state information must be estimated, which poses new challenges due to the almost passive nature of the LIS elements, which have no transmit or receive processing chains. A few works have attempted to tackle this issue by proposing to estimate the cascaded channel in the uplink (UL), and then using channel reciprocity to estimate the equivalent downlink (DL) channel, required for the active and passive beamforming design. These existing methods have adopted a least-squares (LS) criterion for channel estimation \cite{mishra2019channel,nadeem2019intelligent,jensen2019optimal,zheng2019intelligent,you2020intelligent} which, while simple to implement, is also sub-optimal, since it does not take into account the channel statistical features, nor those of the noise. 

In this work, we consider the design of optimal channel estimation strategies for LIS-aided multiple antenna communications, based on the minimum mean squared error (MMSE) criterion. Currently, there is little work on the design of MMSE channel estimators for LIS-aided communications, with the exception of the recent contribution \cite{nadeem2020intelligent}, which considered a configuration with a line-of-sight channel between the BS and LIS, known perfectly to the BS. This strong assumption leads to a linear Gaussian measurement model, and gives a closed form expression for the MMSE channel estimate. 

Here we consider a general scenario in which all channels (BS to LIS, LIS to user equipment (UE), and BS to UE) are subject to Rayleigh fading and are unknown. The optimal MMSE estimator in this case does not admit a closed-form analytical solution, due to the complicated statistical properties of the cascaded channel between the BS, LIS and UE (which does not follow a Gaussian distribution). To deal with this problem, first we derive the best \emph{linear} estimator that minimizes the mean squared error, i.e., the LMMSE, which we show admits a closed-form expression depending on the second-order statistics of the channel and the additive noise. The LMMSE performance depends on the phase shifts employed at the LIS during channel estimation, which may be optimized. Finding the optimum phase shifts is complex however, since it involves solving a non-convex optimization problem. To approximately solve this problem, we present an  algorithm based on the majorization-minimization (MM) principle, which presents an efficient numerical solution. The MM-based algorithm demonstrates the existence of numerous locally-optimal LIS phase-shift matrices. These include the analytical DFT phase-shift matrix that has been shown to be optimal for the case of LS channel estimators \cite{jensen2019optimal}. The DFT-based phase-shift matrix also presents a desirable practical solution for LMMSE, which we show also allows for tractable analytical performance characterization.  

We then move beyond the class of linear filters to present algorithms for approximating the optimal (non-linear) MMSE channel estimator for LIS-aided multiple-antenna communications. To this end, we introduce data-driven deep learning approaches. Specifically, we propose convolutional neural network (CNN)-based channel estimators that approximate the optimal MMSE solution. Our approach is to consider a linear LS channel estimate as a noisy `image' at the neural network input, and then to apply a CNN-based image denoising network to `clean' this image and yield an improved LIS channel estimate.  We consider two CNN-based architectures, Denoising CNN (DnCNN) \cite{zhang2017beyond}
and Fast and Flexible Denoising Network (FFDNet) \cite{zhang2018ffdnet}, that have different properties and performance trade-offs.  
Our numerical simulations show that the CNN-based channel estimators can offer significant performance improvements when compared to the linear estimators, LMMSE and LS. The high performance and reasonably low implementation complexity of these CNN-based channel estimators makes them promising candidates for practical incorporation into future LIS-aided wireless communication systems.


The rest of the paper is organized as follows. Section \ref{systemmodel} presents the system model, defines the channel estimation problem, and reviews the existing LS channel estimator. Section \ref{lmmse} presents our derivation of the LMMSE channel estimator, and develops the MM-based algorithm to optimize the LIS phase-shift matrix. In Section \ref{cnn}, we introduce the CNN-based channel estimators, while Section \ref{numerical} presents numerical results. Finally, some concluding remarks are offered in Section \ref{conclusion}.


{\em Notation}: Matrices and vectors are represented by boldface upper case ($\bm{A}$) and boldface lower case ($\bm{a}$) letters respectively. Hermitian (conjugate transpose), inverse, and trace of a matrix $\bm{A}$ are denoted by $\bm{A}^{H}$, $\bm{A}^{-1}$ and ${\rm tr}(\bm{A})$ respectively. ${\rm diag}(\bm{a})$ with $\bm{a} \in {\mathbb C}^{N}$ returns a $N\times N$ diagonal matrix with $\bm{a}$ on  its diagonal, and ${\rm vec}(\bm{A})$ returns a vector by stacking the columns of $\bm{A}$. ${\mathbb E}[\cdot]$ denotes the expectation operator and $\otimes$  the Kronecker product. $\bm{B} = [\bm{A}]_{i:j, k:l}$ denotes the matrix formed by the elements in the $i-$th to $j-$th rows and $k-$th to $l-$th columns of $\bm{A}$.
$\bm{1}_N \,, \bm{0}_N \in {\mathbb C}^{N}$ represent the vector of all ones and all zeros respectively.

\section{System Model} \label{systemmodel}
\begin{figure}%
    \centering
   {\includegraphics[width=\linewidth]{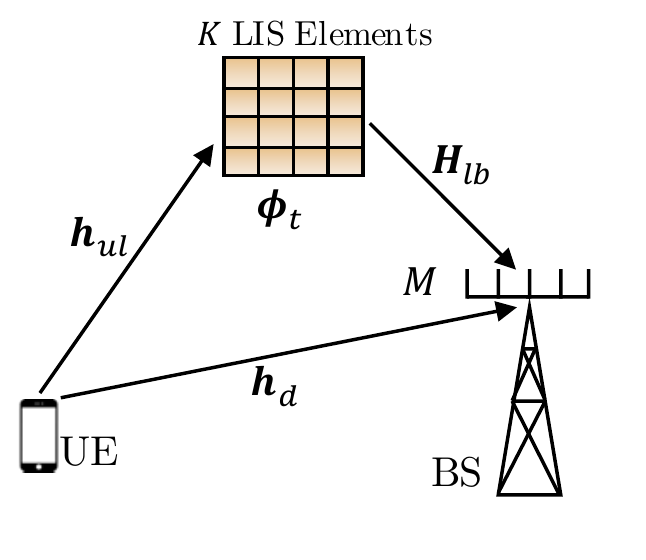} }
   \vspace{-0.5in}
   \caption{Schematic of the LIS-aided MISO communication system.   }
    \label{fig:2.1}%
\end{figure}

We consider a multiple-input single-output (MISO) communication system where a BS with $M$ antennas serves a single antenna UE with the help of an LIS having $K$ passive elements; see Fig. \ref{fig:2.1}. The LIS elements can intelligently control the phase of the incoming electromagnetic wave. Assuming the system operates in time-division duplex mode, the DL channel can be estimated from the UL channel, due to channel reciprocity. The LIS elements are passive with no computing power, hence the UL channel is estimated at the BS. During training, the phase shifts of the LIS elements can be configured to assist the UL channel estimation. The received pilot signal at the BS during the $t$-th training step is given by

\begin{equation}
    \bm{y}_t = (\bm{h}_d + \bm{H}_{lb} {\rm diag}(\bm{\phi}_t)\bm{h}_{ul} )x_t + \bm{n}_t
    \label{e2.1}
\end{equation}
where $\bm{y}_t \in {\mathbb C}^{M}$ is the received signal at the BS, and $x_t \in \mathbb{C}$, $|x_t|=1$ is the transmitted pilot symbol from the UE. Also, $\bm{h}_d \in {\mathbb C}^{M} $ is the direct channel between the UE and BS, $\bm{H}_{lb}\in {\mathbb C}^{M \times K} $ is the channel matrix between the LIS and BS, $\bm{h}_{ul} \in {\mathbb C}^{K} $ is the channel between the UE and LIS, and $\bm{n}_t \sim {\mathcal CN}(0, \sigma^2 \bm{I}_{M})$ represents additive white Gaussian noise (AWGN) at the BS. The vector $\bm{\phi}_t = [e^{j \theta_{t,1}}, \ldots, e^{j \theta_{t,K}} ]^{T} \in {\mathbb C}^{K}$ is the phase shift vector of the LIS, where $ 0 \leq \theta_{t,k} \leq 2\pi, \; k=1,\ldots,K, \; t=1,\ldots,T_p$ is the phase shift of the $k-$th LIS element during the $t$-th training step.  Throughout the paper, we define the training SNR as $\gamma_{{\rm tr}}= 1/\sigma^2 $.

Due to the passive nature of the LIS elements, the cascaded channel $\bm{V} := \bm{H}_{lb} {\rm diag}(\bm{h}_{ul}) =[\bm{v}_1, \bm{v}_2, \ldots ,\bm{v}_K] $ is estimated, rather than separately estimating $\bm{H}_{lb}$ and $\bm{h}_{ul}$. The model (\ref{e2.1}) may be equivalently expressed in terms of $\bm{V}$ as \cite{jensen2019optimal}
 \begin{equation}
    \bm{y}_t = (\bm{h}_d + \bm{V} \bm{\phi}_t )x_t + \bm{n}_t
    \label{e2.2} \; .
 \end{equation}
We assume the channels are subjected to quasi-static Rayleigh fading, which remain constant during the coherence time $T_c > T_p$, where $T_p$ is the pilot duration. The direct channel vector between UE and BS is modelled as \cite{gesbert}
 \begin{equation}
     \bm{h}_d = \bm{R}_{ub}^{1/2} \bm{h}_{wd}
     \label{e2.3}
 \end{equation}
 where $\bm{h}_{wd} \sim {\mathcal CN} (\bm{0}, \bm{I}_K) $, and $\bm{R}_{ub}$ is an $M \times M$ positive definite matrix with unit diagonal entries, representing the spatial correlation matrix at the BS. The channel matrix between the LIS and BS is modelled as \cite{gesbert}
 \begin{equation}
     \bm{H}_{lb} = \bm{R}_{lb}^{1/2} \bm{H}_{wlb}\bm{S}_{lb}^{1/2} \; ,
     \label{e2.4}
 \end{equation}  
where $\bm{H}_{wlb} \sim {\mathcal CN} (\bm{0}, \bm{I}_M \otimes \bm{I}_K)$ while $\bm{R}_{lb}$ and $\bm{S}_{lb} $  are $M \times M$ and $K \times K$ positive definite matrices with unit diagonal entries, representing the spatial correlation matrices at the UE and LIS respectively, and $\bm{h}_{ul} \sim {\mathcal CN} (\bm{0}, \bm{I}_K) $. Collecting all received signals $\bm{y}_t$ from (\ref{e2.1}) across $t=1,2,\ldots, T_p $, it can be compactly written as a linear measurement model \cite[Eq.~8]{jensen2019optimal}
 \begin{equation}
     \bm{y}= \bm{G} \bm{z} + \bm{n} \; ,
     \label{e2.5}
 \end{equation}
  where
  \begin{equation}
      \bm{y}= \begin{bmatrix}
      \bm{y}_1 \\
      \bm{y}_2\\
      \vdots \\
      \bm{y}_{T_p} \\
      \end{bmatrix} \; , \; \; 
      \bm{n}= \begin{bmatrix}
      \bm{n}_1 \\
      \bm{n}_2\\
      \vdots \\
      \bm{n}_{T_p}\\
      \end{bmatrix} \; ,  \; \;
      \bm{z}= 
      \begin{bmatrix}
      \bm{h}_d \\
      \bm{v}_1\\
      \vdots \\
      \bm{v}_K\\
      \end{bmatrix} \; 
      \label{e2.6}
  \end{equation}
represent the observation vector, noise vector and the unknown vector respectively. Note that $\bm{n} \sim {\mathcal CN}(0, \sigma^2 \bm{I}_{MT_p})$, while, from \cite[Eq.~3-5]{jensen2019optimal}, $\bm{G} = \bm{X}\bm{\Psi}$ with 
  $  \bm{X} = {\rm diag}( [x_1 \bm{1}_{M} , \ldots , x_{T_p}  \bm{1}_{M} ]) \; ,  \bm{\Psi}=\bm{\Phi}\otimes \bm{I}_M $
  and 
  
  \begin{equation}
      \bm{\Phi}= \begin{bmatrix}
1 & \phi_{1,1} & \ldots & \phi_{1,K}\\
\vdots & \vdots & \ddots & \vdots\\
1 & \phi_{T_p,1} & \ldots & \phi_{T_p,K}
    \end{bmatrix} \; .
    \label{e2.7}
  \end{equation}

The LIS channel estimation problem requires estimating the vector $\bm{z}$, which contains the direct and cascaded channel, from the linear measurement model (\ref{e2.5}). 

The optimum estimate of $\bm{z}$ that minimizes the mean squared error (MSE) ${\mathbb E}[||\bm{z}-\hat{\bm{z}}||^2 ]$ is the MMSE estimate. It is given by \cite[Eq.~10.5]{kay1993fundamentals}
\begin{equation}
    \hat{\bm{z}}_{{\rm mmse}} = {\mathbb E}[\bm{z}|\bm{y}] \; .
    \label{epos}
\end{equation}
For the LIS channel model, the unknown random vector $\bm{z}$ is not Gaussian, but rather it involves the cascaded channel, which is a product of independent Gaussians. This makes the posterior distribution $p(\bm{z}|\bm{y})$ complicated, and hence, makes it difficult to find $\hat{\bm{z}}_{{\rm mmse}}$ in closed form. Prior works on LIS channel estimation have tackled this problem by instead using the sub-optimal LS channel estimator, reviewed next.

  \subsection{Existing LS Channel Estimation Approaches}
  The LS estimate of the unknown vector $\bm{z}$ in (\ref{e2.5}) is given by \cite{nadeem2019intelligent,mishra2019channel,jensen2019optimal}
  \begin{equation}
      \hat{\bm{z}}_{{\rm ls}} = \underset{\bm{z}}{{\rm arg min}} ||\bm{y} -\bm{G} \bm{z}||_{2}^{2} = \left(\bm{G}^{H}\bm{G} \right)^{-1}\bm{G}^{H}\bm{y}
      \label{e2.8}
  \end{equation}
  which exists when $T_p \geq K+1$. Using (\ref{e2.5}) and (\ref{e2.8}), the LS estimate can be expressed as
  \begin{equation}
     \hat{\bm{z}}_{{ \rm ls}} = \bm{z} + \bm{w}
     \label{e2.9}
  \end{equation}
where $\bm{w} \sim {\mathcal CN}\left(0, \sigma^2 \left(\bm{G}^{H}\bm{G} \right)^{-1}\right)$.  The quality of the LS estimate depends on $\bm{G}$, which in turn depends on the phase shift matrix $\bm{\Phi}$ at the LIS. Different choices of $\bm{\Phi}$ have been proposed. The authors of \cite{jensen2019optimal} have shown that the choice that minimizes the estimation variance per element is given by the first $K+1$ columns of the $T_p \times T_p$ DFT matrix $[\bm{F}_{T_p}]_{t,k} = e^{-j2\pi (t-1)(k-1)/T_p}$.
This choice of $\bm{\Phi}$ makes $\bm{G}^{H}\bm{G} $ a scaled identity matrix, giving the MSE   \cite[Eq.~31]{jensen2019optimal}
\begin{equation}
    {\rm MSE}_{\hat{\bm{z}}_{{\rm ls}}} =  \frac{M(K+1)}{T_p}\sigma^2  \; .
    \label{e2.14}
\end{equation}
Another simpler method that has been proposed for $\bm{\Phi}$ is one that estimates the cascaded channel by sequentially switching on one LIS element at a time \cite{nadeem2019intelligent,yang2019intelligent,mishra2019channel}. This so-called `on-off' method is sub-optimal, producing a higher MSE than the DFT approach \cite[Eq.~17]{jensen2019optimal}. Note that for both the DFT and on-off approaches, the $\bm{\Phi}$ is fixed, and hence it can be set prior to channel estimation commencing.


Unlike the MMSE estimator introduced in (\ref{epos}), the LS channel estimators are known to be sub-optimal since they do not use the prior knowledge of the channel distribution \cite{1597555}. In the following, we will present enhanced channel estimation solutions, designed based on the MMSE criterion.

\section{ LMMSE Channel Estimator} \label{lmmse}

While it is difficult to find the optimum MMSE channel estimator in closed form, the best {\em linear} estimator that minimizes the MSE---the LMMSE estimator---admits a closed form expression, which we present in the following. This estimator, unlike LS, depends on the second order statistics of both the channel and the noise. 

Using the linear measurement model (\ref{e2.5}), the LMMSE estimate of $\bm{z}$ can be expressed as
  \cite[Eq.~12.26]{kay1993fundamentals}
\begin{equation}
    \hat{\bm{z}}_{{\rm lmmse}} = \bm{C}_{\bm{z}\bm{z}}\bm{G}^{H} \left(\bm{G}\bm{C}_{\bm{z}\bm{z}}\bm{G}^{H}+ \sigma^2 \bm{I}_{MT_p} \right)^{-1} \bm{y} \; ,
    \label{e3.1.1}
\end{equation}
where $ \bm{C}_{\bm{z}\bm{z}}= {\mathbb E}[\bm{z}\bm{z}^{H}]$. Let $\bm{H}_{lb}^i$ be the $i-$th column of $\bm{H}_{lb}$ and $h_{ul}^i$ be the $i-$th element of $\bm{h}_{ul}$, then the $i$-th column of $\bm{V}$ is given by 
\begin{equation}
    \bm{v}_i = h_{ul}^i \bm{H}_{lb}^i \;.
    \label{e3.1.2}
\end{equation}
We note that ${\rm vec}\left(\bm{H}_{lb} \right) \sim {\mathcal CN} (\bm{0}, \bm{S}_{lb} \otimes \bm{R}_{lb}) $, and hence $\bm{H}_{lb}^i \sim {\mathcal CN} (\bm{0}, [\bm{S}_{lb} \otimes \bm{R}_{lb}] _{(i-1)M+1 : iM, (i-1)M+1 : iM} ) $. Since $\bm{h}_d$ and $\bm{v}_i $ are independent $\; \forall \; i=1,2,\ldots, K$, $\bm{C}_{\bm{z}\bm{z}}$ is a block diagonal matrix given by
\begin{equation}
    \begin{split}
        \bm{C}_{\bm{z}\bm{z}} &= {\mathbb E} \left\{ \begin{bmatrix}
        \bm{h}_d \\
        \bm{v}_1 \\
        \vdots\\
        \bm{v}_K
        \end{bmatrix}  
        \begin{bmatrix}
        \bm{h}_{d}^{H} & \bm{v}_{1}^{H} & \ldots & \bm{v}_{K}^{H}
        \end{bmatrix}\right\} \\
        &= \begin{bmatrix}
        \bm{R}_{ub} & \bm{0}_{M\times M}& \ldots & \bm{0}_{M\times M} \\
        \bm{0}_{M\times M} & \bm{R}_1 & \ldots & \bm{0}_{M\times M} \\
        \vdots & \vdots & \ddots & \vdots \\
        \bm{0}_{M\times M} & \bm{0}_{M\times M} &\ldots & \bm{R}_K 
        \end{bmatrix}
    \end{split}
    \label{e3.1.3}
\end{equation}
where $\bm{R}_i= [\bm{S}_{lb} \otimes \bm{R}_{lb}] _{(i-1)M+1 : iM, (i-1)M+1 : iM}.$ 

The error covariance matrix for the LMMSE channel estimate is given by \cite[Eq.~12.29]{kay1993fundamentals}, \cite[Eq.~14]{jensen2019optimal}
\begin{equation}
    \bm{C}_{\hat{\bm{z}}_{{\rm lmmse}}} = \left(\bm{C}_{\bm{z}\bm{z}}^{-1} + \frac{\bm{\Phi}^{H}\bm{\Phi} \otimes \bm{I}_M}{\sigma^2} \right)^{-1}
    \label{e3.1.4} \; ,
\end{equation}
and the MSE of the LMMSE estimator is given by ${\rm MSE}_{\hat{\bm{z}}_{{\rm lmmse}}} = {\rm tr} \left(\bm{C}_{\hat{\bm{z}}_{{\rm lmmse}}} \right) $.

Similar to the LS estimators introduced previously, the  performance of the LMMSE estimator depends on the choice of phase-shift matrix $\bm{\Phi}$.  The optimum phase-shift matrix can be obtained by solving the following optimization problem
\begin{equation}
    \begin{split}
    & \underset{ \bm{\Phi}} {{\rm min}} \quad {\rm tr}\left( \left(\bm{C}_{\bm{z}\bm{z}}^{-1} + \frac{\bm{\Phi}^{H}\bm{\Phi} \otimes \bm{I}_M}{\sigma^2} \right)^{-1} \right)\\ &
    {\rm s.t} \quad \quad \bm{\Phi} \quad {\rm satisfies} \, (\ref{e2.7}) ,\\
    &  \quad \quad \; \; |\phi_{t,k}| = 1, \; t=1,\ldots T_p, \;  k=1,\ldots,K \;.
    \end{split}
    \label{e3.1.5}
\end{equation}
This optimization problem is non-convex due to the unit modulus constraints, and obtaining an analytical solution is difficult. Similar problems also arise when designing optimal pilot designs for MIMO channel estimation schemes \cite{kotecha2004transmit}. In the following we will present an algorithm, employing a MM computational approach, to efficiently solve the optimization problem in (\ref{e3.1.5}).

\subsection{Majorization-Minimization (MM) Based Phase Shift Optimization}
MM refers to a class of  algorithms that solve difficult optimization problems by solving a series of simpler problems which often admit closed-form solutions. MM successively forms a surrogate function or majorizer of the objective function at the current iterate, and then optimizes the surrogate function. These algorithms have been successfully applied to approximately solve difficult non-convex optimization problems in wireless communications and signal processing \cite{huang2019reconfigurable,sun2016majorization,wang2016design}. Here we present an MM-based algorithm to solve the non-convex problem in (\ref{e3.1.5}). Our approach is inspired from the prior work in \cite{wang2016design}, which designed unit-modulus pilot signals for MIMO channel estimation.

We use the minimum pilot duration $T_p=K+1$ and rewrite the objective function of (\ref{e3.1.5}) as follows
\begin{equation}
    {\rm MSE}\left(\bm{\Phi} \right) = {\rm tr}\left(\bm{C}_{\hat{\bm{z}}_{{\rm lmmse}}} \right)
    \label{emm1}
\end{equation}
where 
\begin{equation}
    \bm{C}_{\hat{\bm{z}}_{{\rm lmmse}}} = \left( \bm{C}_{\bm{z}\bm{z}}^{-1} + \left(\bm{\Phi} \otimes \bm{I}_M \right)^{H} \bm{W}^{-1} \left(\bm{\Phi} \otimes \bm{I}_M \right)  \right)^{-1} \;,
    \label{emm2}
\end{equation}
with $\bm{W} = \sigma^{2} \bm{I}_{M(K+1)}$.
Denoting $ \Tilde{\bm{\Phi}} = \bm{\Phi} \otimes \bm{I}_M$ and using \cite[Eq.~23]{wang2016design}, $\bm{C}_{\hat{\bm{z}}_{{\rm lmmse}}} $ can be equivalently expressed as
\begin{equation}
  \bm{C}_{\hat{\bm{z}}_{{\rm lmmse}}} = \bm{C}_{\bm{z}\bm{z}}- \bm{C}_{\bm{z}\bm{z}}\Tilde{\bm{\Phi}}^{H} \left( \Tilde{\bm{\Phi}}\bm{C}_{\bm{z}\bm{z}} \Tilde{\bm{\Phi}}^{H} + \bm{W} \right)^{-1}\Tilde{\bm{\Phi}}\bm{C}_{\bm{z}\bm{z}}\; .
  \label{emm3}
\end{equation}
Hence, the objective function becomes
\begin{equation}
 {\rm MSE}\left(\bm{\Phi} \right) = {\rm tr} \left(\bm{C}_{\bm{z}\bm{z}}- \bm{C}_{\bm{z}\bm{z}}\Tilde{\bm{\Phi}}^{H}\bm{P}^{-1}\Tilde{\bm{\Phi}}\bm{C}_{\bm{z}\bm{z}} \right)   \;
 \label{emm4}
\end{equation}
where $\bm{P} = \left( \Tilde{\bm{\Phi}}\bm{C}_{\bm{z}\bm{z}} \Tilde{\bm{\Phi}}^{H} + \bm{W} \right) $.  This objective function is similar to an objective function considered in \cite[Eq.~40]{wang2016design}, with the main technical differences lying in the differing constraints of our optimization variable $\bm{\Phi}$, and the different ordering of the matrices in the Kronecker product operation in the definition of $\Tilde{\bm{\Phi}} $. Despite these differences, we may apply a similar approach to \cite{wang2016design} to find a surrogate function or majorizer function of ${\rm MSE}\left(\bm{\Phi} \right) $. This function $g_{\rm MSE}$ must satisfy the following
\begin{align}
    & (i) \; g_{{\rm MSE}} \left(\bm{\Phi},\bm{\Phi}_{t} \right) \geq {\rm MSE}\left(\bm{\Phi} \right) \; \forall \; \bm{\Phi} \in {\rm dom}\left(\bm{\Phi} \right) \label{emm5a} \\
    & (ii) \; g_{{\rm MSE}} \left(\bm{\Phi}_{t},\bm{\Phi}_{t} \right) ={\rm MSE}\left(\bm{\Phi}_t \right)
    \label{emm5}
\end{align}
where ${\rm dom}\left(\bm{\Phi} \right) $ contains all $\bm{\Phi}$ that satisfy $ (\ref{e2.7}) $ and $ |\phi_{t,k}| = 1, \; t=1,\ldots K+1, \;  k=1,\ldots,K$. To find the surrogate function we need the following lemma.

\begin{lemma} \label{lemm1}
The function ${\rm tr}\left(\bm{X}^{H}\bm{Y}^{-1}\bm{X} \right)$, with $\bm{Y} \succ 0$, can be lower-bounded as \cite[Ex.~4]{sun2016majorization}
\begin{equation}
\begin{split}
   {\rm tr}\left(\bm{X}^{H}\bm{Y}^{-1}\bm{X} \right) \geq  2 {\rm Re} \{ {\rm tr} (\bm{X}_{t}^{H} \bm{Y}_{t}^{-1}\bm{X})\} - {\rm tr}(\bm{Y}_{t}^{-1}\bm{X}_{t}\bm{X}_{t}^{H}\bm{Y}_{t}^{-1}\bm{Y}) \\ + {\rm const.}
   \end{split}
   \label{emm6}
\end{equation}
with equality at $(\bm{X},\bm{Y})= (\bm{X}_t,\bm{Y}_t)$.
\end{lemma}
\begin{IEEEproof}
The proof  mirrors that of \cite[Ex.~4]{sun2016majorization}, with  adaptations to apply for complex matrices, rather than real matrices.

\end{IEEEproof}
Using the result of Lemma \ref{lemm1} in (\ref{emm4}), a surrogate function satisfying the properties (\ref{emm5a}) and (\ref{emm5}) can be obtained as
\begin{align}
g_{{\rm MSE}} \left(\bm{\Phi},\bm{\Phi}_{t} \right) &= {\rm MSE}\left(\bm{\Phi}_t \right) + {\rm tr}(\bm{A}_t\bm{A}_{t}^{H} \Tilde{\bm{\Phi}}\bm{C}_{\bm{z}\bm{z}}\Tilde{\bm{\Phi}}^{H}) \nonumber \\
 & \hspace*{1cm}  -2{\rm Re}\{ {\rm tr}(\bm{C}_{\bm{z}\bm{z}}\bm{A}_{t}^{H}\Tilde{\bm{\Phi}}) \} \; ,
 \label{emm7}
\end{align}
where $\Tilde{\bm{\Phi}}_t = \bm{\Phi}_t \otimes \bm{I}_{M} $ and $\bm{A}_t = \left( \Tilde{\bm{\Phi}}_t\bm{C}_{\bm{z}\bm{z}} \Tilde{\bm{\Phi}}_{t}^{H} + \bm{W} \right)^{-1}\Tilde{\bm{\Phi}}_t\bm{C}_{\bm{z}\bm{z}}$. The non-convex problem in (\ref{e3.1.5}) can now be approximately solved by iteratively solving the following  optimization problem
\begin{equation}
    \begin{split}
    & \underset{ \bm{\Phi}} {{\rm min}} \quad g_{{\rm MSE}} \left(\bm{\Phi},\bm{\Phi}_{t} \right)\\ &
    {\rm s.t} \quad \quad \bm{\Phi} \quad {\rm satisfies} \, (\ref{e2.7}) ,\\
    &  \quad \quad \; \; |\phi_{t,k}| = 1, \; t=1,\ldots K+1, \;  k=1,\ldots,K 
    \end{split}
    \label{emm8}
\end{equation}
However, to efficiently solve (\ref{emm8}), it is helpful to apply a second majorization to upper bound $g_{{\rm MSE}} \left(\bm{\Phi},\bm{\Phi}_{t} \right)$. For this, we use the following lemma.
\begin{lemma} \label{lemm2}
\cite[Lemma~4]{wang2016design}
For Hermitian matrices $\bm{M} \in {\mathbb C}^{n\times n}$, $\bm{Z} \in {\mathbb C}^{m\times m}$, and any $\bm{X} \in {\mathbb C}^{m\times n}$, the function ${\rm tr}(\bm{Z}\bm{X}\bm{M}\bm{X}^{H})$ is majorized by $-2{\rm Re}\{{\rm tr}\left( \left(\lambda\bm{X}_t - \bm{Z}\bm{X}_t\bm{M} \right)^{H}\bm{X}\right) \} + \lambda||\bm{X}||_{F}^{2} + {\rm const.} $, where $\lambda \bm{I} \succeq \bm{M}^{T}\otimes\bm{Z}$.
\end{lemma}


Using Lemma \ref{lemm2} with $\bm{Z}=\bm{A}_t \bm{A}_{t}^{H}, \bm{X}=\Tilde{\bm{\Phi}}$ and $\bm{M}=\bm{C}_{\bm{z}\bm{z}}$, $ g_{{\rm MSE}} \left(\bm{\Phi},\bm{\Phi}_{t} \right) $ can be upper-bounded as

\begin{equation}
    \begin{split}
        g_{{\rm MSE}} \left(\bm{\Phi},\bm{\Phi}_{t} \right) \leq -2{\rm Re}\{{\rm tr}\left( \left(\lambda_t \Tilde{\bm{\Phi}}_t - \bm{A}_t\bm{A}_{t}^{H}\Tilde{\bm{\Phi}}_t\bm{C}_{\bm{z}\bm{z}} \right)^{H}\Tilde{\bm{\Phi}}\right) \} \\
        -2{\rm Re}\{ {\rm tr}(\bm{C}_{\bm{z}\bm{z}}\bm{A}_{t}^{H}\Tilde{\bm{\Phi}}) \}
        + \lambda_t||\Tilde{\bm{\Phi}}||_{F}^{2} + {\rm const.}
    \end{split}
    \label{emm9}
\end{equation}
From the constraints in (\ref{emm8}), we note that $||\Tilde{\bm{\Phi}}||_{F}^{2} = {\rm const.}$ Thus, the equivalent problem to be solved at each iteration is given by
\begin{equation}
    \begin{split}
    & \underset{ \bm{\Phi}} {{\rm min}} \quad -2{\rm Re}\{{\rm tr}(\bm{B}^{H}\Tilde{\bm{\Phi}}) \}\\ &
    {\rm s.t} \quad \quad \Tilde{\bm{\Phi}} = \bm{\Phi} \otimes \bm{I}_{M} \\ &
    \quad  \quad \quad \bm{\Phi} \quad {\rm satisfies} \, (\ref{e2.7}) ,\\
    &  \quad \quad \; \; |\phi_{t,k}| = 1, \; t=1,\ldots K+1, \;  k=1,\ldots,K 
    \end{split}
    \label{emm10}
\end{equation}
where $\bm{B}=\lambda_t \Tilde{\bm{\Phi}}_t - \bm{A}_t\bm{A}_{t}^{H}\Tilde{\bm{\Phi}}_t \bm{C}_{\bm{z}\bm{z}} + \bm{A}_t \bm{C}_{\bm{z}\bm{z}}$, and $\lambda_t \bm{I} \succeq \bm{C}_{\bm{z}\bm{z}}^{T} \otimes \left(\bm{A}_t\bm{A}_{t}^{H} \right)$. The tightest upperbound of $\lambda_t$ is given by $ \lambda_t = \lambda_{{\rm max}} \left( \bm{C}_{\bm{z}\bm{z}}^{T} \otimes \left(\bm{A}_t\bm{A}_{t}^{H} \right) \right)$. Since the computational cost of computing the largest eigenvalue can be large due to the large matrix dimensions, we use the following approximation for $\lambda_t$ \cite[Eq.~61]{wang2016design}
\begin{equation}
  \lambda_t = ||\bm{C}_{\bm{z}\bm{z}} ||_{1} ||\bm{A}_t\bm{A}_{t}^{H}||_{1} \; ,
  \label{emm11}
\end{equation}
where $||\cdot||_{1}$ denotes the maximum absolute column sum matrix norm \cite{horn2012matrix}. Moreover, since $\Tilde{\bm{\Phi}} = \bm{\Phi} \otimes \bm{I}_{M}$, it can be shown that 
\begin{equation}
  {\rm tr}(\bm{B}^{H}\Tilde{\bm{\Phi}}) ={\rm tr}\left( \left(\sum_{i=1}^{M} \bm{B}_i \right)^{H} \bm{\Phi} \right)
  \label{emm12}
\end{equation}
where 
\begin{equation}
    \bm{B}_i = \bm{B}_{i:M:i+KM,i:M:i+KM}
    \label{emm13}
\end{equation}
denotes the submatrix of $\bm{B}$ extracted from the $K+1$ rows and columns of $\bm{B}$ with indices $[i, i+M,\ldots,i+MK] $. Denote $\Tilde{\bm{B}}= \sum_{i=1}^{M} \bm{B}_i $, the equivalent optimization problem to be solved at each iteration is given by 
\begin{equation}
  \begin{split}
    & \underset{ \bm{\Phi}} {{\rm min}} \quad -2{\rm Re}\{{\rm tr}(\Tilde{\bm{B}}^{H}\bm{\Phi}) \}\\ &
    {\rm s.t}
    \quad  \quad  \bm{\Phi} \quad {\rm satisfies} \, (\ref{e2.7}) ,\\
    &  \quad \quad \; \; |\phi_{t,k}| = 1, \; t=1,\ldots K+1, \;  k=1,\ldots,K 
    \end{split}
    \label{emm14}   
\end{equation}
The MM update $\bm{\Phi}_{t+1}$ at each iteration can be obtained by solving the following equivalent optimization problem
\begin{equation}
  \begin{split}    & \underset{ \bm{\Phi}} {{\rm min}} \quad ||\Tilde{\bm{B}}-\bm{\Phi}||_{F}^{2}\\ &
    {\rm s.t}
    \quad  \quad  \bm{\Phi} \quad {\rm satisfies} \, (\ref{e2.7}) ,\\
    &  \quad \quad \; \; |\phi_{t,k}| = 1, \; t=1,\ldots K+1, \;  k=1,\ldots,K 
    \end{split}
    \label{emm15}   
\end{equation}
which can be solved by projecting $\Tilde{\bm{B}} $ onto the constraint set as follows
\begin{align}
    [\bm{\Phi}_{t+1}]_{:,2:K+1} &= \exp\{j\phase{[\Tilde{\bm{B}}]_{:,2:K+1} }\} \nonumber \\
    [\bm{\Phi}_{t+1}]_{:,1} &= \bm{1}_{K+1}
    \label{emm16}
\end{align}
where $[\bm{A}]_{:,i:j}$ denotes elements in all the rows from the $i$-th to the $j$-th column of $\bm{A}$. The overall MM based algorithm is summarized in Algorithm \ref{Algo1}.


It has been shown that MM based algorithms converge to a stationary point for bounded objective functions \cite{song2015optimization}. Using similar arguments as in \cite[Thm.~1]{wang2016design}, it can be shown that the generated sequence of points $\bm{\Phi}_t, t=0,1,\ldots$ monotonically decreases the objective function ${\rm MSE}\left(\bm{\Phi} \right)$, and the algorithm converges to a stationary point.

\begin{algorithm}\label{Algo1}
\DontPrintSemicolon
\SetAlgoLined
\KwInput{$\epsilon, \bm{W}, \bm{C}_{\bm{z}\bm{z}} $}
\KwOutput{$\bm{\Phi}$ }
 Set $t=0$ and initialize $[\bm{\Phi}_{0}]_{i,j}= e^{j \theta_{i,j}}$ with \ $\theta_{i,j} \sim {\rm Unif} [0,2\pi] \; \forall \, i= 1,\ldots,K+1, j= 2,\ldots, K+1 $ and $[\bm{\Phi}_{0}]_{:,1} = \bm{1}_{K+1}. \; \Tilde{\bm{\Phi}}_0 = \bm{\Phi}_0 \otimes \bm{I}_{M} $  \;
 $ {\rm MSE}_{0} =  {\rm tr}\left(\left( \bm{C}_{\bm{z}\bm{z}}^{-1} + \Tilde{\bm{\Phi}}_{0}^{H} \bm{W}^{-1} \Tilde{\bm{\Phi}}_{0}  \right)^{-1} \right) $\\
 \Repeat{${\rm MSE}_{t-1}-{\rm MSE}_{t}\leq \epsilon $}
 {
  $\Tilde{\bm{\Phi}}_t = \bm{\Phi}_t \otimes \bm{I}_{M}$\;
  $\bm{A}_t = \left( \Tilde{\bm{\Phi}}_t\bm{C}_{\bm{z}\bm{z}} \Tilde{\bm{\Phi}}_{t}^{H} + \bm{W} \right)^{-1}\Tilde{\bm{\Phi}}_t\bm{C}_{\bm{z}\bm{z}}$ \;
  $ \lambda_t = ||\bm{C}_{\bm{z}\bm{z}} ||_{1} ||\bm{A}_t\bm{A}_{t}^{H}||_{1}$\;
 $\bm{B}=\lambda_t \Tilde{\bm{\Phi}}_t - \bm{A}_t\bm{A}_{t}^{H}\Tilde{\bm{\Phi}}_t \bm{C}_{\bm{z}\bm{z}} + \bm{A}_t \bm{C}_{\bm{z}\bm{z}}$ \;
 $\Tilde{\bm{B}}= \sum_{i=1}^{M} \bm{B}_{i:M:i+KM,i:M:i+KM}$\;
 $[\bm{\Phi}_{t+1}]_{:,2:K+1} = \exp\{j\phase{[\Tilde{\bm{B}}]_{:,2:K+1} }\}$ \;
 $[\bm{\Phi}_{t+1}]_{:,1} = \bm{1}_{K+1}$\;
 $\Tilde{\bm{\Phi}}_{t+1} = \bm{\Phi}_{t+1} \otimes \bm{I}_{M}$\;
 $ {\rm MSE}_{t+1} =  {\rm tr}\left(\left( \bm{C}_{\bm{z}\bm{z}}^{-1} + \Tilde{\bm{\Phi}}_{t+1}^{H} \bm{W}^{-1} \Tilde{\bm{\Phi}}_{t+1}  \right)^{-1} \right) $\;
  $t=t+1$\;
 }
 \caption{MM-based algorithm for solving (\ref{e3.1.5}).}
\end{algorithm}

\begin{figure}[h] 
\centering
\includegraphics[width=0.5\textwidth]{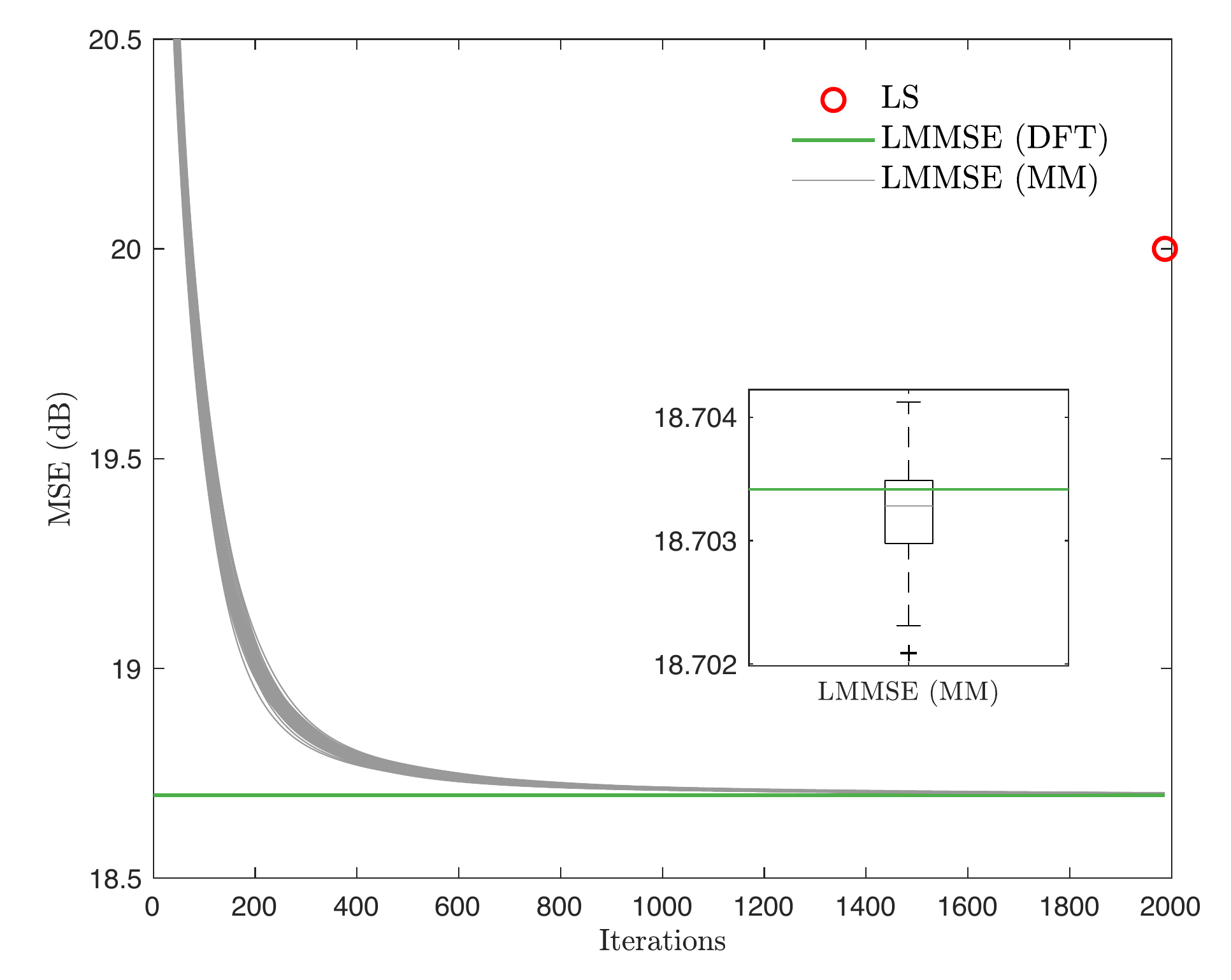}%
\caption{The plots show the MSE of $\hat{\bm{z}}$ obtained from the LMMSE channel estimator as the iterations of the MM algorithm increase, with $M=10, K=50$ at $\gamma_{{\rm tr}}= -10$ dB for $100$ different random initialization points. The MSE of the LS and the LMMSE channel estimator with $\bm{\Phi}$ equal to the DFT matrix are also shown for comparison. }
\label{Fig2}
\end{figure}

We demonstrate the performance of Algorithm \ref{Algo1} for designing the phase shift matrix $\bm{\Phi}$ by considering an example simulation scenario with $M=10, K=50$, exponential spatial correlation matrices with $[\bm{R}_{ub}]_{i,j}= [\bm{R}_{lb}]_{i,j} = 0.6^{|i-j|}$ , $[\bm{S}_{lb}]_{i,j}= 0.6^{|i-j|}$, and $\gamma_{{\rm tr}}= -10$ dB. 
Fig. \ref{Fig2} plots the MSE of $\hat{\bm{z}}$ obtained from the LMMSE channel estimator as the iterations of Algorithm \ref{Algo1} increase. The plot shows trajectories of the MM algorithm for $100$ different random initialization points. For comparison it also shows the MSE obtained from the LS estimator, and the LMMSE channel estimator with $\bm{\Phi}$ chosen equal to the DFT matrix. It is observed that the phase-shift matrices initialized with random phases have poor MSE (even worse than the LS estimator). However, as the iterations of the MM algorithm increase, the MSE monotonically decreases for the different initialization points, and all solutions converge to a stationary point. Further, we observe that the DFT matrix is also a stationary point that attains almost the same MSE as the solutions produced by Algorithm \ref{Algo1}, with the added advantage of being independent of the noise variance and channel statistics. The box-plot shows the distribution of the MSE of the different local optima produced by Algorithm \ref{Algo1} with the 100 different random initialization points. We observe that the MSE of local optima are very close to each other and also to the DFT based solution with the difference being of the order $10^{-3}$ dB.

\subsection{MSE Analysis for DFT Based Design}

Here we further analyse the performance of the LMMSE channel estimator, considering the DFT-based phase shift matrix.  As we have shown, the DFT matrix achieves a local optima in terms of MSE performance that matches closely with the numerical MM-based solutions, despite being very simple (i.e., not requiring numerical iterations), while it is also more amenable to further analytical study.

With the DFT-based phase shift matrix, the MSE of the LMMSE channel estimator can be written as 
\begin{equation}
    {\rm MSE}_{\hat{\bm{z}}_{{\rm lmmse}}}  =  {\rm tr}\left(\left(\bm{C}_{\bm{z}\bm{z}}^{-1} + \frac{T_p\bm{I}_{M(K+1)}}{\sigma^2} \right)^{-1} \right)
    \label{e3.1.6} \; ,
\end{equation}
which can be equivalently expressed as
\begin{equation}
    {\rm MSE}_{\hat{\bm{z}}_{{\rm lmmse}}}= \sum_{i=1}^{M} \left( \frac{1}{\frac{1}{\lambda_i\left(\bm{R}_{ub} \right) }+ \frac{T_p}{\sigma^2} } + \frac{K}{\frac{1}{\lambda_i\left(\bm{R}_{lb} \right) }+ \frac{T_p}{\sigma^2} } \right) \;,
    \label{e3.1.8}
\end{equation}
where $\lambda_{i}\left( \bm{R}_{ub} \right), \lambda_{i}\left( \bm{R}_{lb} \right) $ are the eigenvalues of $ \bm{R}_{ub}$ and 
$ \bm{R}_{lb}$ respectively. From \cite[Corr.~2.5]{palomar2007mimo}, the function $f(\bm{x})= \sum_{i=1}^{n} \frac{1}{\frac{1}{x_i} + c}$ is Schur-concave in $\bm{x}=[x_1,\ldots,x_n]$, since $\frac{1}{\frac{1}{x_i} + c}$ is a concave function in $x_i$. Thus, the MSE in (\ref{e3.1.8}) is Schur-concave in both $\bm{\lambda}\left(\bm{R}_{ub} \right)$ and $\bm{\lambda}\left(\bm{R}_{lb} \right)$. Further, due to the trace constraint on the covariance matrices, the vector of all ones (corresponding to the eigenvalues of the identity matrix) is majorized by the eigenvalue vector of a covariance matrix whose eigenvalues are spread away from one\cite[Lemma~2.2]{palomar2007mimo}, i.e.,
\begin{equation}
  \bm{\lambda}\left( \bm{R}_{ub}\right)   \succ [1,\ldots,1] \,, \; {\rm and} \; \bm{\lambda}\left( \bm{R}_{lb}\right)   \succ [1,\ldots,1]\;.
   \label{e3.1.9}
\end{equation}
Thus, the MSE is maximized when $\bm{R}_{ub}=\bm{R}_{lb} =\bm{I}_{M} $ and the maximum MSE is given by
\begin{equation}
   {\rm MSE}_{\hat{\bm{z}}_{{\rm lmmse}}} ^{{\rm max}} = \frac{M(K+1)}{1+ \frac{T_p}{\sigma^2}} \;.
   \label{e3.1.10}
\end{equation}
When comparing two covariance matrices $\bm{R}_x$ and $\bm{R}_y$, it is said that $\bm{R}_x$ is `more correlated' than $\bm{R}_y$ if $\bm{\lambda}\left(\bm{R}_x \right) \succ \bm{\lambda}\left(\bm{R}_y \right)$ \cite[Sec.~ 4.1.2]{jorswieck2007majorization}. Thus, as the degree of correlation of $\bm{R}_{ub}$ and 
$ \bm{R}_{lb}$ increases, the MSE of the LMMSE channel estimate decreases. Note also that, even for the MSE maximizing scenario (\ref{e3.1.10}), it holds that ${\rm MSE}_{\hat{\bm{z}}_{{\rm lmmse}}} ^{{\rm max}} < {\rm MSE}_{\hat{\bm{z}}_{{\rm ls}}} $, with ${\rm MSE}_{\hat{\bm{z}}_{{\rm ls}}}$ the MSE of the LS estimate, given by (\ref{e2.14}). Hence, the MSE of the LMMSE channel estimate is lower than that of the LS estimate for all SNR ranges, and also for all spatial correlation profiles.

\begin{figure}[htp] 
\centering
\includegraphics[width=0.5\textwidth]{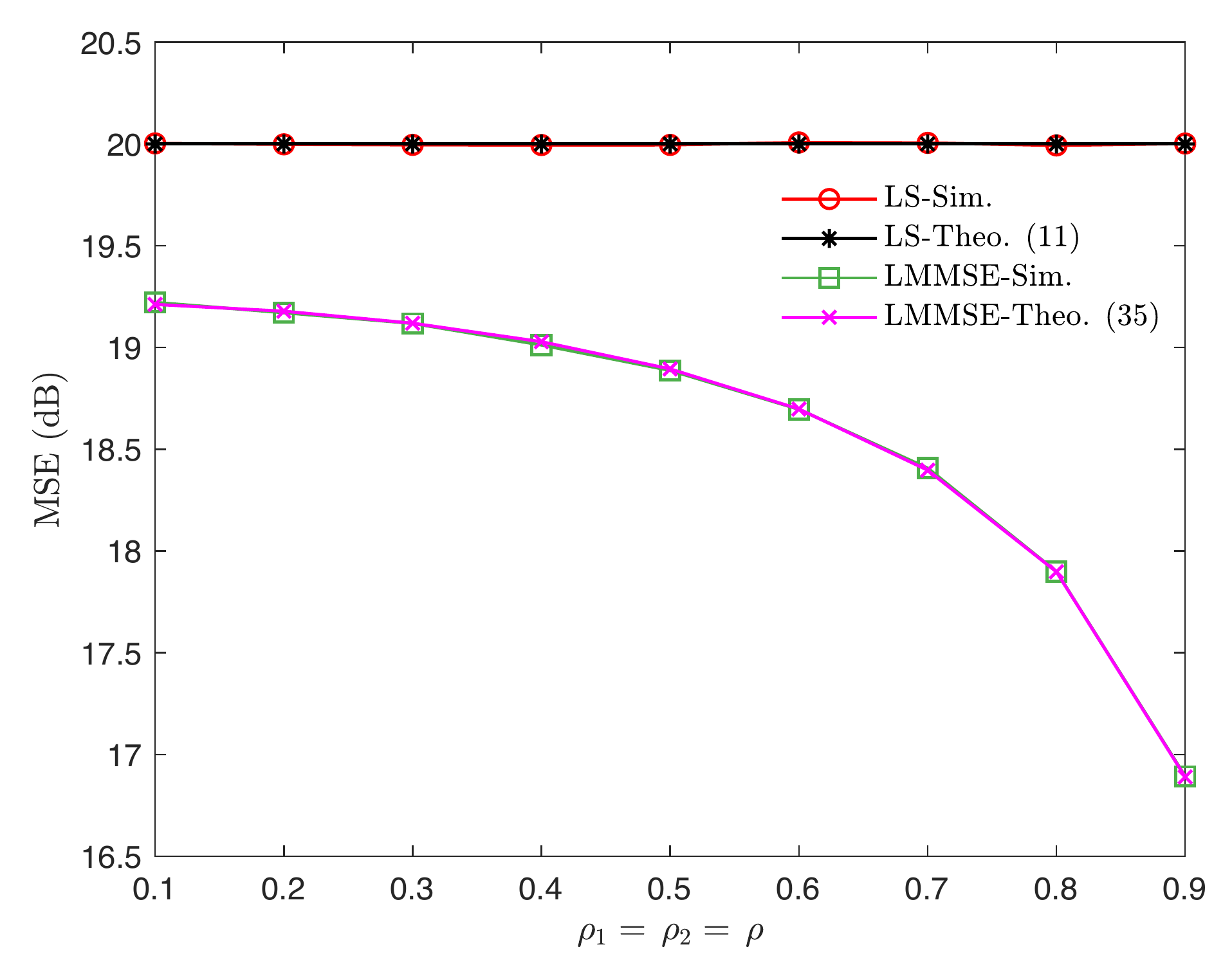}%
\caption{ The plots compare the MSE of the LS and LMMSE channel estimate of the combined direct and cascaded channel $\bm{z}$, as the correlation coefficients $\rho_1, \rho_2$ increase. Results shown for $M=10, K=50$ and $\gamma_{{\rm tr}}= -10$ dB.}
\label{fig3:a}
\end{figure}

We present a comparison of the MSE performance of the LMMSE and LS channel estimates in Fig. \ref{fig3:a}, considering the same simulation scenario as introduced in the previous subsection. Fig. \ref{fig3:a} shows the theoretical and the simulated MSE performance of the LS and LMMSE estimate of the combined direct and cascaded channel $\bm{z}$ at an SNR of $ -10$ dB, as a function of the spatial correlation coefficients. The theoretical MSE for the LS and LMMSE estimates are calculated using (\ref{e2.14}) and (\ref{e3.1.8}) respectively. It is observed that while the MSE of the LS estimate remains constant, regardless of the spatial correlation, the MSE of the LMMSE channel estimate decreases as the channels become more correlated, in line with the theoretical analysis presented above.

We may further analytically quantify the MSE improvement of the LMMSE estimate over the simple LS estimate at high and low SNR conditions. 
Using a Taylor series expansion at high SNR (i.e., for $\sigma^2 \to 0$), the MSE of the LMMSE estimate in (\ref{e3.1.8}) can be expressed as
\begin{align}
  {\rm MSE}_{\hat{\bm{z}}_{{\rm lmmse}}} &= \nonumber \\
  & \hspace*{-1cm} {\rm MSE}_{\hat{\bm{z}}_{{\rm ls}}} - \left(\frac{\sigma^2}{T_p} \right)^2 \left( {\rm tr}\left( \bm{R}_{ub}^{-1} \right) + K{\rm tr}\left( \bm{R}_{lb}^{-1} \right) \right) + o\left(\sigma^4\right) \; .
  \label{e3.1.11}
\end{align}
This shows that, to leading order, the performance of the LMMSE estimator approaches that of the LS estimator as the SNR grows very large. However, for large but finite SNR, there is a  performance improvement of the LMMSE approach determined by the trace of the inverse spatial covariance matrices. The effect of $\bm{R}_{lb} $ is relatively higher than that of $\bm{R}_{ub} $ on the MSE improvement, since $\bm{R}_{lb} $ appears in the LMMSE estimate for each of the $K$ effective cascaded SIMO channels from the UE to the BS, via each LIS element. 

Considering now the case of low SNR (i.e., for $\sigma^2 \to \infty$), the MSE in (\ref{e3.1.8}) admits
\begin{equation}
    {\rm MSE}_{\hat{\bm{z}}_{{\rm lmmse}}} = M(K+1) - \frac{T_p}{\sigma^2} \left( {\rm tr}\left( \bm{R}_{ub}^{2} \right) + K{\rm tr}\left( \bm{R}_{lb}^{2} \right) \right) + o\left(\frac{1}{\sigma^2}\right) \, .
    \label{e3.1.12}
\end{equation}
Importantly, the MSE remains bounded as $\sigma^2 \to \infty$. This is in contrast to the LS estimate, as evident from (\ref{e2.14}), which shows that ${\rm MSE}_{\hat{\bm{z}}_{{\rm ls}}} \to \infty$ as $\sigma^2 \to \infty$.  Hence, while the LMMSE estimator yields improved performance over LS for all (finite) SNR values, the gain is particularly significant in the low SNR regime. 



The asymptotic expansions can be further simplified for the exponential spatial correlation model, for which $[\bm{R}_{ub}]_{i,j}= \rho_{1}^{|i-j|}, \; [\bm{R}_{lb}]_{i,j} = \rho_{2}^{|i-j|}$ with $ 0 < \rho_{1}, \rho_{2} < 1$. Specifically, the MSE at high SNR (\ref{e3.1.11}) simplifies to
\begin{equation}
\begin{split}
   {\rm MSE}_{\hat{\bm{z}}_{{\rm lmmse}}} ={\rm MSE}_{\hat{\bm{z}}_{{\rm ls}}}-   \left(\frac{\sigma^2}{T_p} \right)^2 \Bigg[ \frac{M+(M-2)\rho_{1}^2}{1-\rho_{1}^2}  \\ 
   + \frac{K\left(M+(M-2)\rho_{2}^2\right)}{1-\rho_{2}^2} \Bigg] + o\left(\sigma^4\right) \; ,
\end{split}
\label{e3.1.13}
\end{equation}
while the MSE at low SNR (\ref{e3.1.12}) simplifies to
\begin{equation}
\begin{split}
    {\rm MSE}_{\hat{\bm{z}}_{{\rm lmmse}}}  = M(K+1) - \frac{T_p}{\sigma^2} \Bigg[ \frac{M(1-\rho_{1}^4)-2\rho_{1}^2(1-\rho_{1}^{2M})}{\left(1-\rho_{1}^2\right)^2} \\
    + \frac{K\left(M(1-\rho_{2}^4)-2\rho_{2}^2\left(1-\rho_{2}^{2M}\right) \right)}{\left(1-\rho_{2}^2\right)^2} \Bigg] + o\left(\frac{1}{\sigma^2}\right) \, .
\end{split}
    \label{e3.1.14}
\end{equation}
Here, the high SNR expression follows after applying the following property of the exponential correlation matrix \cite[Eq.~30]{mallik2018exponential},
\begin{equation}
    {\rm tr} \left( \bm{R}^{-1} \right) =\frac{M+(M-2) \rho^{2}}{1-\rho^{2}}\;,
    \label{e3.1.15}
\end{equation}
whereas the low SNR expression follows upon noting 
\begin{equation}
    {\rm tr} \left( \bm{R}^2 \right)= \|\bm{R} \|_{F}^{2} = \sum_{i,j} \rho^{2|i-j|} = \frac{M(1-\rho^4)-2\rho^2\left(1-\rho^{2M} \right)}{\left(1-\rho^2\right)^2}\;
    \label{e3.1.16}
\end{equation}
for an $M\times M$ exponential correlation matrix $[\bm{R}]_{i,j}= \rho^{|i-j|}$.

An advantage of the expressions in (\ref{e3.1.13}) and (\ref{e3.1.14}) is that they quantify the MSE of the LMMSE estimate explicitly in terms of the correlation coefficients $\rho_1, \rho_2$ at high and low SNR respectively. From these expressions it is clear that the MSE of the LMMSE channel estimate decreases monotonically as the correlation coefficients $\rho_1, \rho_2$ increase (with the effect of $\rho_2$ being relatively larger than that of $\rho_1$). This is in line with the general conclusions made previously based on majorization arguments.

\section{ Deep Learning Approaches to MMSE Channel Estimation} \label{cnn}

Despite yielding performance improvements over LS, the proposed LMMSE estimator is only optimal within the class of linear estimators.  Beyond the class of linear estimators, the optimal MMSE estimator is difficult to determine analytically, due to the difficulty in evaluating the posterior mean in (\ref{epos}). To address this issue, we now propose data-driven non-linear channel estimators that approximate the optimal MMSE solution using deep neural networks. 

In general, our approach is to introduce CNN-based estimators that take as input the sub-optimal LS estimate and produce as output an improved channel estimate that removes noise from the LS estimate, and approximates the optimal MMSE solution. Our approach is motivated by the universal function approximation property of CNNs, which states that a deep CNN can be used to approximate any continuous function to an arbitrary degree of accuracy when the depth of the network is large enough and given enough training samples \cite{zhou2020universality}.
\begin{figure*}[h]%
    \centering
  {\includegraphics[width=\textwidth]{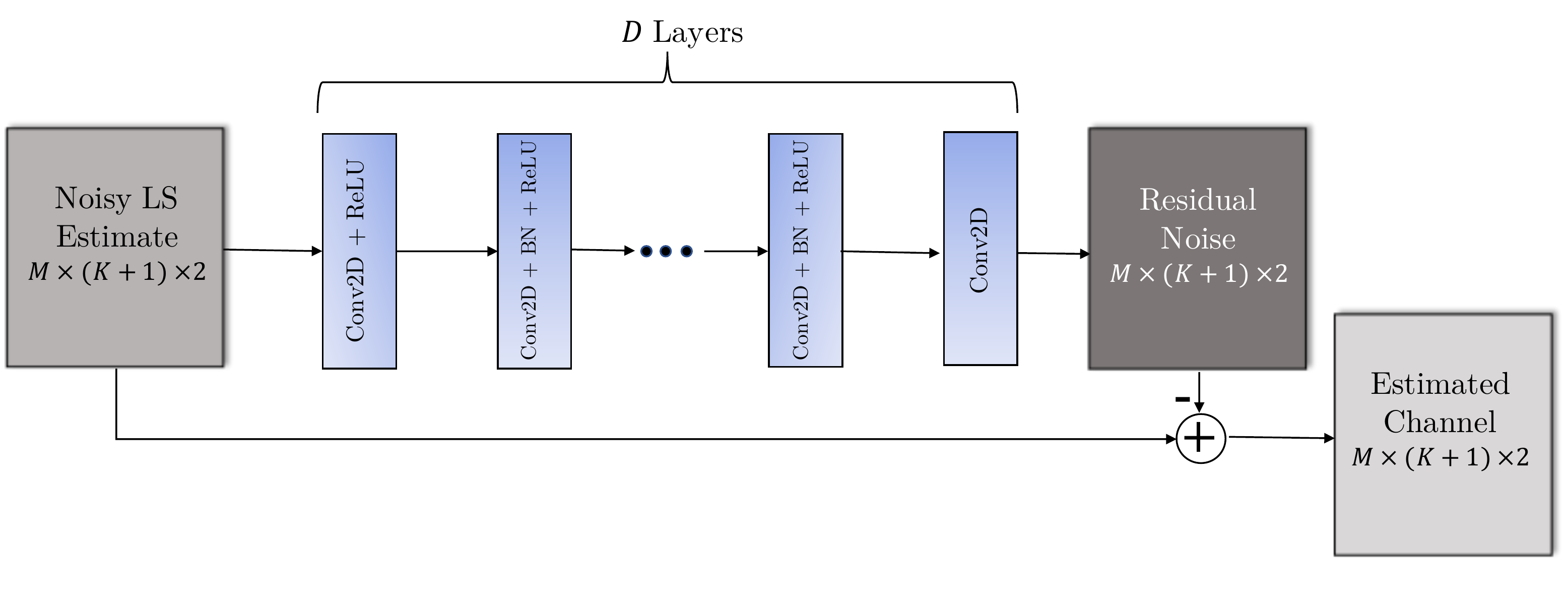} }
  \vspace{-0.5in}
  \caption{Network architecture of the DnCNN-based channel estimator. }
    \label{fig:3.1}%
\end{figure*}

 \begin{figure*}[!b]%
\centering
{\includegraphics[width=\textwidth]{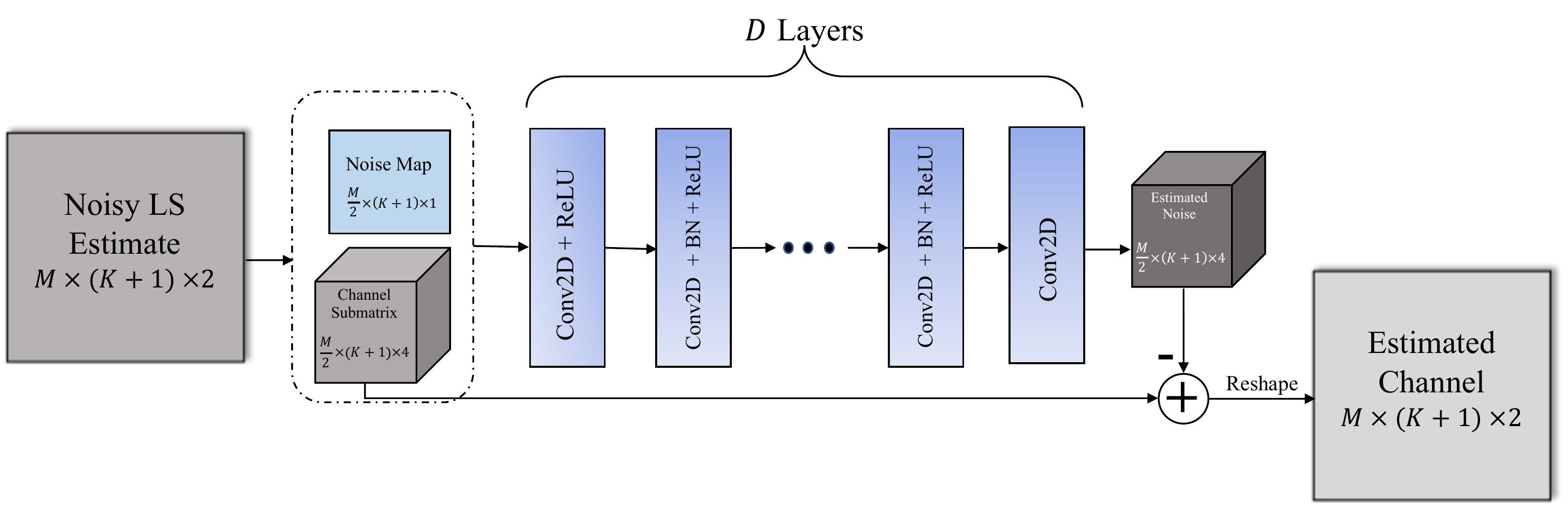} }
\vspace{-0.5in}
\caption{Network architecture of the FFDNet-based channel estimator. }
\label{fig:3.2}%
\end{figure*}

We use the LS channel estimator with DFT-based phase-shift matrix ${\bf \Phi}$ as the input to our CNN. (One could also use the LMMSE estimate as input, however we choose the LS estimate due to its simplicity, and since it does not require  knowledge of the noise variance nor the channel correlation matrices.)  With this choice, it is convenient to reshape (\ref{e2.9}) in matrix form as 
    \begin{equation}
     \hat{\bm{Z}}_{{\rm ls}} = \bm{Z} + \bm{\zeta} \; ,
     \label{e3.2.1}
  \end{equation}
  \vspace{-0.15in}
  where 
  \begin{equation}
      \begin{split}
         \bm{Z}= [\bm{h}_d, \bm{v}_1, \bm{v}_2,\ldots, \bm{v}_K]\; ,\\
         \hat{\bm{Z}}_{{\rm ls}}= [\hat{\bm{h}_d}_{{\rm ls}}, \hat{\bm{v}_1}_{{\rm ls}}, \hat{\bm{v}_2}_{{\rm ls}},\ldots, \hat{\bm{v}_K}_{{\rm ls}}]
      \end{split}
      \label{3.2.2}
  \end{equation}
and $\bm{\zeta}$ is the additive noise whose elements are independent and identically distributed (i.i.d) as ${\mathcal CN}(0,\frac{\sigma^2}{T_p})$ \cite[Eq.~31]{jensen2019optimal}. In this form, the LS estimate $\hat{\bm{Z}}_{{\rm ls}}$ may be seen as a noisy version of the unknown `image' $\bm{Z}$. 

CNN-based deep learning image denoising algorithms, which have been well studied by the image processing community, may be applied to learn the mapping from noisy images to clean images \cite{zhang2017beyond,zhang2018ffdnet}, and therefore to filter out the image noise. Based on these ideas, we present two different approaches to learn the MMSE channel estimator for the LIS system. First, we present an approach based on the DnCNN architecture \cite{zhang2017beyond}, which does not use knowledge of the noise variance. Then, we present an approach based on the FFDNet architecture \cite{zhang2018ffdnet}, which utilizes the noise variance information to further improve channel estimation performance, particularly at low SNR. In terms of speed, FFDNet has a faster inference time, due to its lower model complexity when compared to DnCNN.

\subsection{DnCNN Network Architecture} 
The input to the DnCNN is the LS channel estimate with dimension $M\times (K+1)$. Since deep learning architectures are naturally designed to deal with real-valued data, we reshape the LS estimate into a tensor of size $M\times (K+1) \times 2$, where the real and imaginary parts of $\hat{\bm{Z}}_{{\rm ls}}$ are separated and placed as two separate channels, similar to the RGB channels in image data. The DnCNN network architecture is shown in Fig. \ref{fig:3.1}. The input tensor is processed by a number of convolutional layers to produce the output. The depth of the network is $D$, with each layer using filters of size $3\times3$, along with zero padding of the input feature matrix such that the output contains multiple feature maps of the same dimension, i.e. $M\times (K+1)$. The number of feature maps $N_f$ is fixed for all the layers except the last layer, which has $2$ feature maps to match the dimension of the input tensor. The first layer performs a convolution operation (Conv2D),  then applies the ReLU non-linear activation function. The middle $D-2$ layers perform Conv2D, apply batch normalization (BN), and then apply the ReLU activation function. BN helps to 
speed up the training process and also leads to better generalization performance of the network \cite{ioffe2015batch}. Finally, the last layer applies the Conv2D operation, but does not apply the non-linear activation function to avoid filtering out negative values. 

With DnCNN, instead of directly learning the mapping from the noisy LS channel estimate to the noise-free channel matrix, we learn the noise map first, and then subtract it from the input noisy LS estimate to get the cleaned channel matrix. It has been shown in the original DnCNN paper \cite{zhang2017beyond} that learning the noise map gives better image denoising performance across different signal-to-noise ratios (SNRs) compared with directly learning the denoised image. Moreover, the residual learning strategy where the input layer is directly connected to the output layer by a skip connection is much easier to train, and provides better generalization performance on the test dataset\cite{he2016deep}.

The operation ${\cal F}$ of the DnCNN-based channel estimator can be described as 
  \begin{equation}
      \bm{\hat{Z}}_D = {\mathcal F}(\hat{\bm{Z}}_{{\rm ls}};
      \bm{\Theta})
      \label{e3.2.3}
  \end{equation}
where $\bm{\hat{Z}}_{{\rm ls}}$  is the noisy LS channel matrix, $\bm{\hat{Z}}_D$ is the  estimated channel matrix output by the DnCNN, and $\bm{\Theta}$ is the neural network parameters that are optimized during the initial training process. The DnCNN is trained by minimizing the MSE loss over the training set, comprising $N_{{\rm tr}}$ labelled data  $\{\bm{Z}^{i}, \hat{\bm{Z}}_{{\rm ls}}^{i} \}_{i=1}^{N_{{\rm tr}}}$, generated from an ensemble of channel realizations. The MSE loss is defined as
  \begin{equation}
      {\mathcal L}\left(\bm{\Theta} \right)= \frac{1}{N_{{\rm tr}}} \sum_{i=1}^{N_{{\rm tr}}} ||\bm{Z}^{i} -{\mathcal F}(\hat{\bm{Z}}_{{\rm ls}}^{i}; \bm{\Theta})||_{F}^{2} \; .
      \label{eloss}
  \end{equation}
The training of DnCNN can be carried out offline by using simulated channel realizations. Once the training is complete, the DnCNN may be employed at the BS for UL channel estimation.

 \subsection{FFDNet Architecture}
The network architecture of DnCNN does not utilize knowledge of the additive noise variance. The authors in \cite{zhang2018ffdnet} have proposed a CNN-based image denoising architecture, FFDNet, which provides better denoising performance by utilizing the noise variance information. We now present the architecture of FFDNet that utilizes the noise variance information to further improve the channel estimation performance.

Similar to the DnCNN, the input to the FFDNet is the noisy LS channel estimate of size $M \times (K+1) \times 2$, after separating the real and imaginary parts. In order to reduce the inference time and improve the efficiency of the channel estimator, the input tensor is further reshaped into a size of $\frac{M}{2} \times (K+1) \times 4$ \cite{zhang2018ffdnet}. Further, the noise map $\bm{N}$ of shape $\frac{M}{2}\times(K+1)$ having each entry set to $\frac{\sigma}{\sqrt{2T_p}}$ (the standard deviation of the additive noise in the LS estimate (\ref{e3.2.1})) is concatenated with the reshaped LS channel matrix such that the input to the FFDNet based channel estimator is a tensor of shape $\frac{M}{2} \times (K+1) \times 5$. This provides flexibility to the network to handle different noise levels. Fig. \ref{fig:3.2} shows the network architecture of the FFDNet based channel estimator. Similar to DnCNN, the input tensor is processed by a number of convolutional layers. The network has a depth of $D$ layers where the first layer uses the Conv2D + ReLU operation, the subsequent $D-1$ layers use the Conv2D + BN + ReLU operation, and finally the last layer uses only the Conv2D operation to output the estimated noise map. The filter size is fixed to $3\times3$. The number of feature maps in all the convolutional layers is fixed to $N_f$ except the last layer which has $4$ feature maps to match the shape of the input channel submatrices. All the convolutional layers use `same' padding in the Conv2D operation so that the output dimension is fixed to $ \frac{M}{2} \times (K+1) \times 4$. Similar to DnCNN, we use a residual learning strategy and learn the mapping of the residual noise from the noisy LS estimate and subtract the residual noise from the input to get the denoised channel submatrices which are then reshaped to obtain the estimated channel output from the FFDNet. The FFDNet-based channel estimator is formulated as 
  \begin{equation}
      \bm{\hat{Z}}_F = {\mathcal G}(\hat{\bm{Z}}_{{\rm ls}}, \bm{N}; \bm{\Theta})
      \label{e3.2.4}
  \end{equation}
where $\hat{\bm{Z}}_{{\rm ls}}$ is the noisy input LS channel matrix, $\bm{N}$ is the noise map which depends on the variance of the additive noise, $\bm{\hat{Z}}_F$ is the estimated channel matrix output of FFDNet, and $\bm{\Theta}$ is the neural network parameter matrix which is optimized during the initial training process. Unlike the DnCNN (\ref{e3.2.3}), FFDNet can adapt to different noise levels through the noise map $\bm{N}$. The training dataset for FFDNet contains $N_{{\rm tr}}$ labelled data points $\{\bm{Z}^{i}, \bm{N}^i, \hat{\bm{Z}}_{{\rm ls}}^{i} \}_{i=1}^{N_{{\rm tr}}}$. Similar to DnCNN, FFDNet is also trained offline by minimizing the MSE loss function 
  \begin{equation}
      {\mathcal L}\left(\bm{\Theta} \right)= \frac{1}{N_{{\rm tr}}} \sum_{i=1}^{N_{{\rm tr}}} ||\bm{Z}^{i} -{\mathcal G}(\hat{\bm{Z}}_{{\rm ls}}^{i},\bm{N}^i; \bm{\Theta})||_{F}^{2} \; 
      \label{lossffdnet}
  \end{equation}
 using simulated channel realizations. After the offline training of FFDNet it can be employed at the BS for UL channel estimation.

 \section{Performance Results} \label{numerical}
 
\subsection{Channel Estimation Performance}
In this section, we present the performance of the DnCNN and FFDNet based channel estimators for LIS assisted MISO communications, and compare their performance with the LMMSE (with DFT phase-shift matrix) and LS estimators.  We consider simulation scenarios with $M=10$ BS antennas and $K=50$ or $K=10$ LIS elements. The pilot duration is set to $T_p = K+1$. As for the previous figures, we use the exponential correlation model for the spatial correlation matrices,
  \begin{equation}
      [\bm{R}_{ub}]_{i,j} = \rho_{1}^{|i-j|} \;,  [\bm{R}_{lb}]_{i,j} = \rho_{2}^{|i-j|} \; , [\bm{S}_{lb}]_{i,j} = \rho_{3}^{|i-j|}
      \label{e4.1}
  \end{equation}
where $0 < \rho_1, \rho_2, \rho_3 <1$. 

Training of the DnCNN is done at a constant SNR of $\gamma_{{\rm tr}}=0 \, {\rm dB}$. The dataset for training FFDNet contains samples from $3$ different SNR levels $\gamma_{{\rm tr}}=-5\, {\rm dB}  , 0\, {\rm dB}$ and $5\, {\rm dB}$. This changes the input noise map accordingly and should help FFDNet to learn the residual noise better than DnCNN. The training, validation and test datasets contain $16000$, $8000$, $6000$ samples respectively. The network is trained using the well-known Adam algorithm \cite{kingma1412adam} with the default parameters $\beta_1= 0.9, \beta_2=0.999$, and with learning rate $0.001$ and mini-batch size $100$. The network is trained until the validation loss remains unchanged for $5$ consecutive epochs.  
\begin{figure*}[h] 
\centering
\subfigure[Direct Channel ($K=50$)]{%
\includegraphics[width=0.5\textwidth]{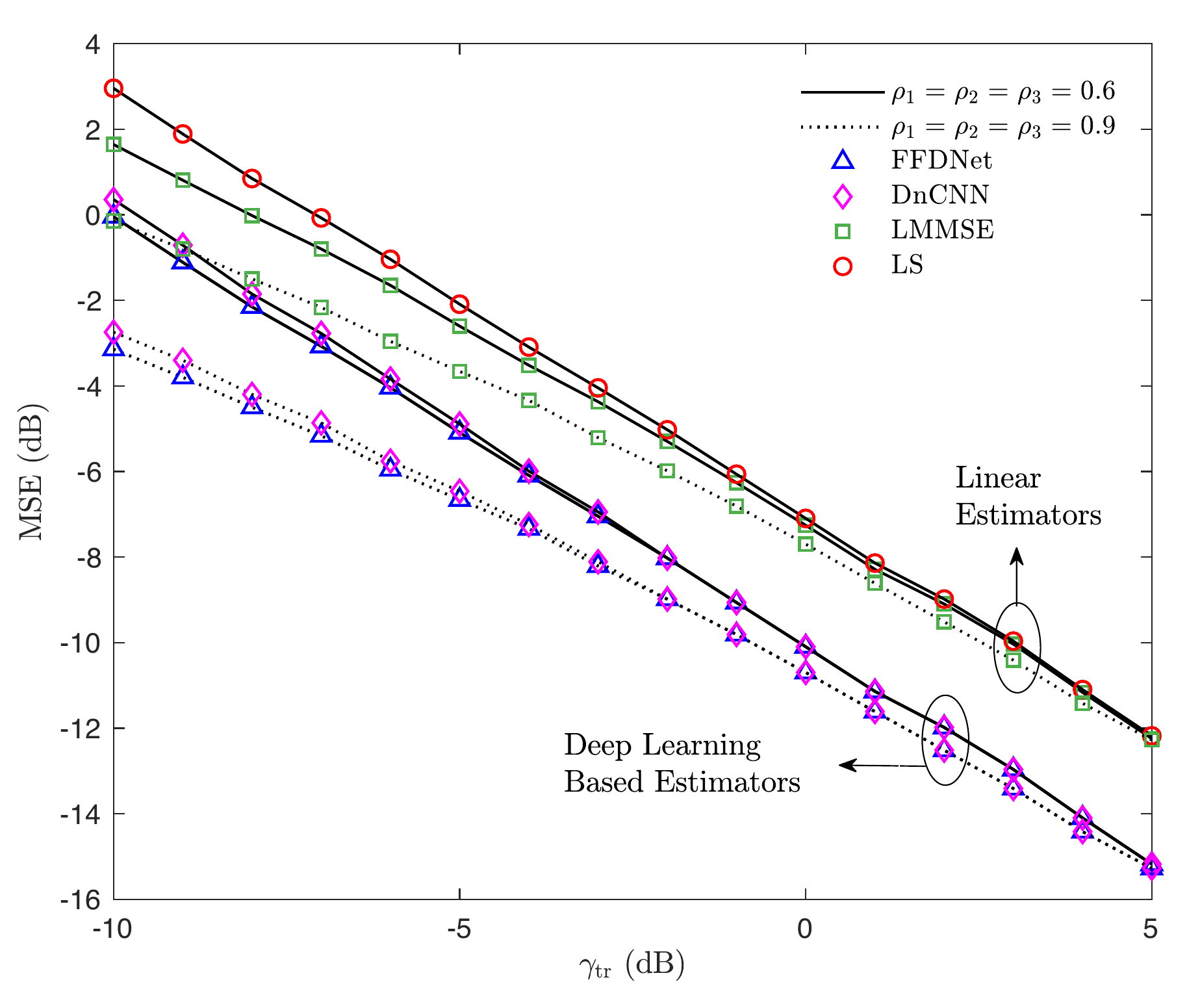}%
\label{fig4.2:a}%
}\hfil
\subfigure[Cascaded Channel ($K=50$)]{%
\includegraphics[width=0.5\textwidth]{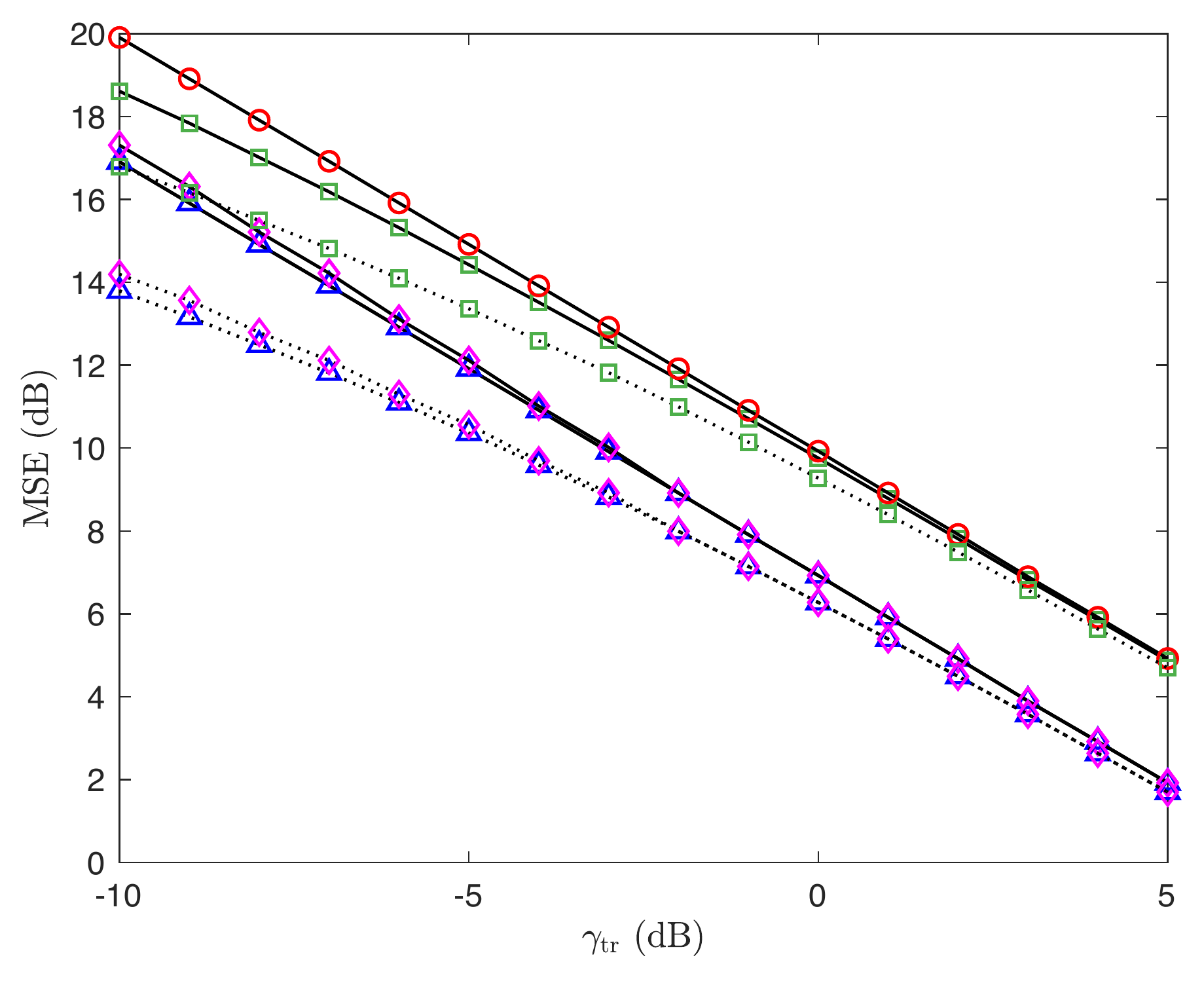}%
\label{fig4.2:b}%
}
\subfigure[Direct Channel ($K=10$)]{%
\includegraphics[width=0.5\textwidth]{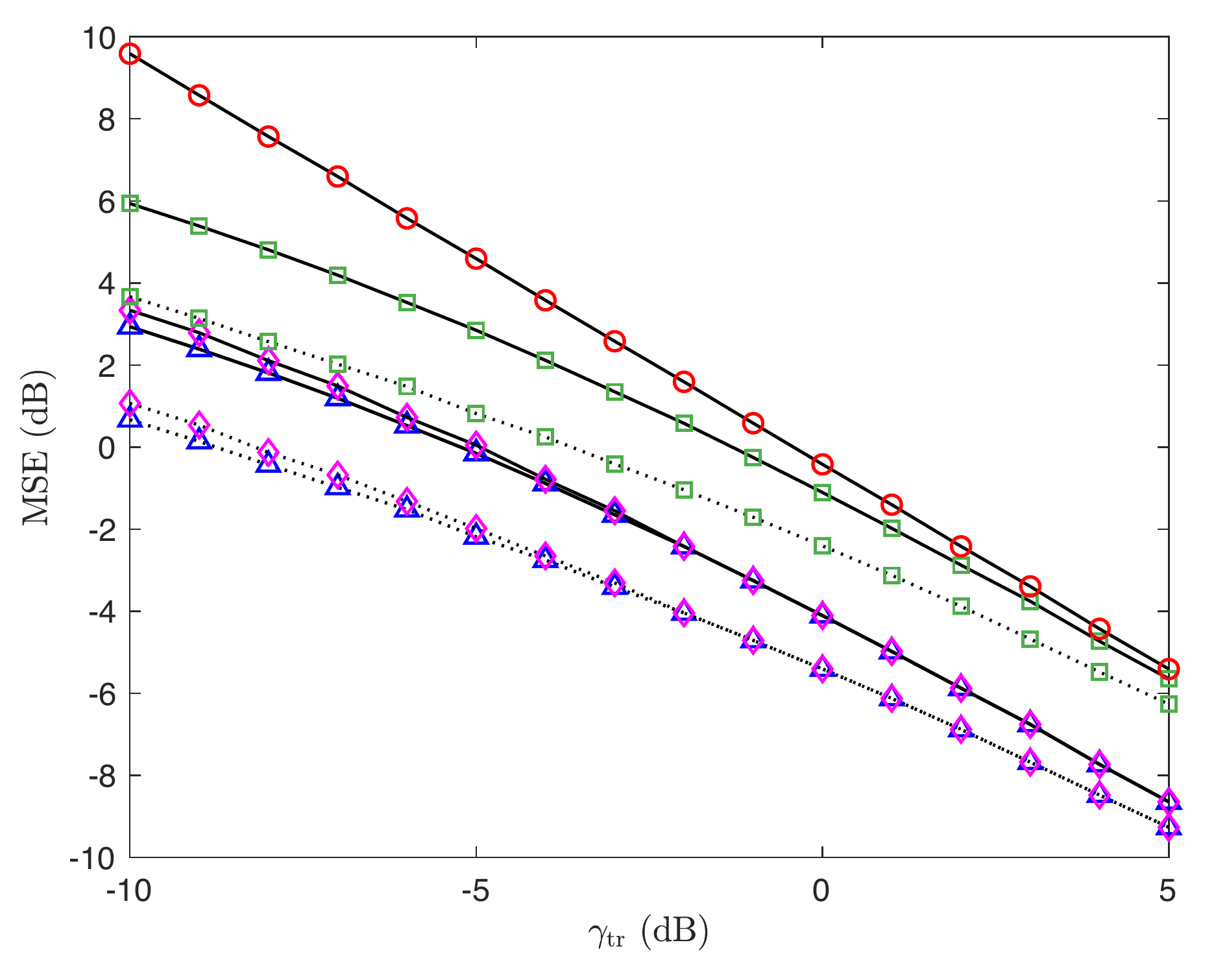}%
\label{fig4.2:c}%
    }\hfil
\subfigure[Cascaded Channel ($K=10$)]{%
\includegraphics[width=0.5\textwidth]{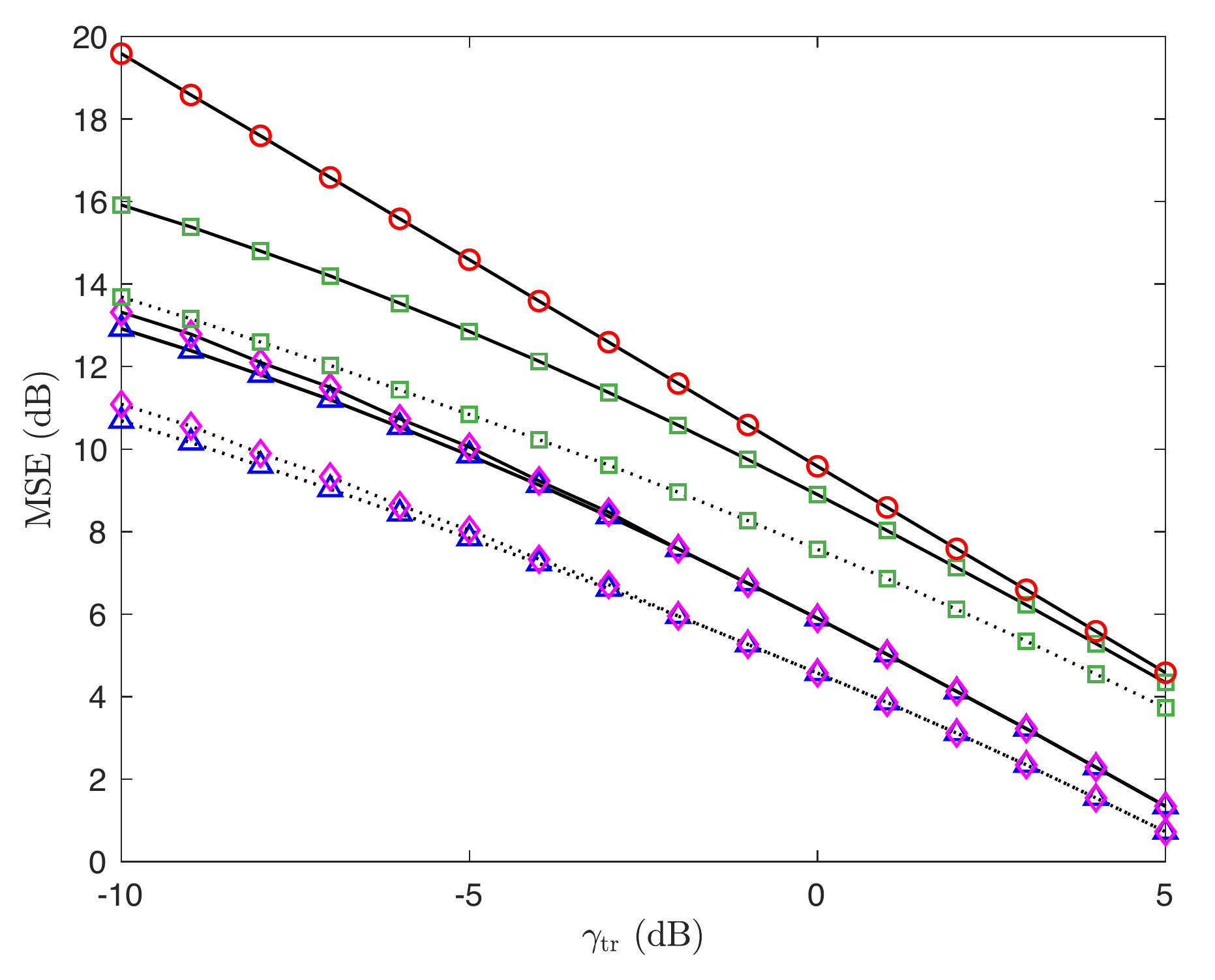}%
\label{fig4.2:d}%
}
\caption{The plots show the MSE performance of LS, LMMSE, DnCNN and FFDNet for the direct channel and the cascaded channel with $M=10$ and two different values of $K=10,50$. The MSE performance is shown for two spatial correlation scenarios. The hyperparameters are chosen as $D=8, N_f=4$, for both DnCNN and FFDNet. 
}
\label{fig4.2}
\end{figure*}

The MSE performance comparison of the different estimators is shown in Fig. \ref{fig4.2} for the direct channel $\bm{h}_d$ and the cascaded channel $\bm{V}$.  For DnCNN and FFDNet, results are shown for $D=8$ and $N_f=4$. It is observed that the performance of DnCNN and FFDNet is superior to that of both linear estimators, LMMSE and LS. The superiority over LMMSE demonstrates the advantages of applying non-linear estimation, allowing the CNN-based estimators to capture the (non-Gaussian) distributional properties of the channel, whereas the LMMSE utilizes only the second order statistics. Note that while the direct channel $\bm{h}_d$ has a Gaussian distribution, the estimate of $\bm{h}_d$ obtained from the LMMSE estimate is not optimal since the received signal at the BS comprises a superposition of signals from the direct and the cascaded channels. Hence, while LMMSE yields the optimal MMSE solution for Gaussian channels, for the MISO configuration with LIS, using a non-linear channel estimator leads to a better MSE performance than the LMMSE estimator even for the direct channel. It is also observed that the MSE performance of DnCNN and FFDNet based channel estimators improve as the channels become more highly correlated, similar to the case of LMMSE, as shown previously. Among the deep learning based solutions, the FFDNet-based channel estimator performs slightly better than the DnCNN-based channel estimator at low SNR due to its ability to adapt to different SNR values. 

\begin{table*}[h]
   \centering
   \caption{Performance of the DnCNN-and FFDNet-based channel estimators for different choices of hyperparameters, $N_f$ and $D$ for $M=10, K=50$ at $ \gamma_{{\rm tr}}= 5\, {\rm dB}$.}
\begin{tabular}{|c|l|c|c|c|c|c|c|c|c|c|}
\hline
\multicolumn{2}{|c|}{\multirow{2}{*}{}}       & $N_f$            & \multicolumn{4}{c|}{4}        & \multicolumn{4}{c|}{8}        \\ \cline{3-11} 
\multicolumn{2}{|c|}{}                        & $D$             & 4     & 6     & 8     & 10    & 4     & 6     & 8     & 10    \\ \hline
\multicolumn{2}{|c|}{\multirow{2}{*}{FFDNet}} & Direct Channel MSE (dB)       & -14.44 & -14.77  & -15.07 & -15.01 & -14.54 & -14.53 & -14.85 & -14.96 \\ \cline{3-11} 
\multicolumn{2}{|c|}{}                        & Cascaded Channel MSE (dB) & 2.47 & 2.15 & 1.91 & 1.97 & 3.39 & 2.38 & 2.08 & 2.2 \\
\cline{3-11} 
\multicolumn{2}{|c|}{}                        & Run Time (ms) & 115.7 & 122.6 & 130.5 & 140.5 & 118.7 & 127.6 & 142.6 & 156.6 \\ \hline
\multicolumn{2}{|c|}{\multirow{2}{*}{DnCNN}} & Direct Channel MSE (dB)       & -14.28 & -14.5  & -15.05 & -15.08 & -13.60 & -14.22 & -14.76 & -14.82 \\ \cline{3-11} 
\multicolumn{2}{|c|}{}                        & Cascaded Channel MSE (dB) & 2.51 & 2.25 & 1.98 & 2.15 & 3.52 & 2.45 & 2.15 & 2.26 \\
\cline{3-11} 
\multicolumn{2}{|c|}{}                        & Run Time (ms) & 171.5 & 192.4 & 217.4 & 232.2 & 187.6 & 218.9 & 232.9 & 257.3 \\ \hline
\end{tabular}
\label{table:1}
\end{table*}


Next, consider the effect of varying $K$. For the direct channel $\bm{h}_d$, while the number of channel elements to be estimated does not depend on $K$, the pilot duration $T_p=K+1$  is longer for the case $K=50$ compared with $K=10$. This leads to a smaller MSE, as observed by comparing Fig. \ref{fig4.2:a} and Fig. \ref{fig4.2:c}. For the LMMSE estimator, this trend is also evident from the theoretical MSE expression (\ref{e3.1.8}). Specifically, the MSE of the direct channel is captured by the first term in (\ref{e3.1.8}), which is clearly decreasing in $K$ when the pilot length is set to $T_p=K+1$. The same is true for the LS estimate, for which the MSE of the direct channel is given by  \cite[~Eq.(30)]{jensen2019optimal} $ \frac{M\sigma^2}{K+1}$. 

For the cascaded channel $\bm{V}$, from Figures \ref{fig4.2:b} and \ref{fig4.2:d}, the opposite trend is observed, with the MSE for the $K=50$ case generally being larger than that for the case $K=10$, most notably at low training SNR. Unlike for the direct channel, for the cascaded channel the number of elements to estimate increases with $K$, making the effect of increasing $K$ on the MSE performance less obvious.  On the one hand, the increased pilot length for larger $K$ reduces the estimation error per channel element, while on the other hand, the errors accumulate across a larger set of estimated channel coefficients. Our results suggest that the aggregation of channel estimation errors has a more significant effect, particularly in the low SNR regime. This is seen also from the theoretical MSE expression for the LMMSE estimate (\ref{e3.1.8}), where for low training SNR (i.e., large $\sigma^2$) the MSE of the cascaded channel, reflected by the second term in (\ref{e3.1.8}), is seen to monotonically and almost linearly increase with $K$. As the training SNR is increased, a monotonic increase of MSE in $K$ is still observed from (\ref{e3.1.8}), albeit now at a slower rate. It is also noteworthy that the MSE for the LS estimator, given theoretically by $\sigma^2 M \frac{K}{K+1}$ \cite[~Eq.(30)]{jensen2019optimal}, becomes almost independent of $K$ for $K \gg 1$, a trend that is also evident from Figures \ref{fig4.2:b} and \ref{fig4.2:d}. 

In interpreting the effect of changing $K$, we add the caveat that so far we have focused only on the MSE performance. The fact that increasing $K$ also leads to an increase in the pilot length can reduce spectral efficiency, by allowing less time for data transmission.  This system-level issue will be analyzed in the following subsection. Prior to conducting this analysis however, we explore the effect of the different hyperparameters of the neural network based estimators, in terms of performance and complexity. 


There are two main hyperparameters for the DnCNN and FFDNet based channel estimators, $D$ and $N_f$. Table \ref{table:1} shows the MSE of the direct and the cascaded channel for two different feature maps $N_f=4,8$ and for different depths of the network $D=4,6,8,10$, at a fixed training SNR of $ \gamma_{{\rm tr}}= 5$ dB with $M=10, K=50$. It also shows the average CPU runtime (in ms) of DnCNN and FFDNet for inference on a test data set of size 1000. It is observed that the MSE performance is not very sensitive to the choice of $N_f$ and $D$. Also, the difference in runtimes are quite modest, with an increase in average runtime observed as $N_f$ and $D$ increase, due to the higher model complexity.   Generally, the setting $N_f=4$ and $D=8$ provides a good trade-off between MSE performance and average runtime for both FFDNet and DnCNN. Further note that the average runtime of FFDNet is lower than that of DnCNN, in accordance with the original FFDNet based image denoising paper \cite{zhang2018ffdnet}.

\subsection{System-Level Performance: Downlink Data Transmission}
Next  we demonstrate the system level performance gains in terms of achievable rate when the proposed channel estimates are employed for DL data transmission. During the data transmission phase, the received signal at the UE is given by \cite[Eq.~3]{mishra2019channel}
\begin{equation}
    y_{u}= \sqrt{P_{tx}}(\bm{h}_{d}^{T} + \bm{\phi}_{d}^{T}\bm{V}^T)\bm{w}x_d + n_d
    \label{ed1}
\end{equation}
where $\bm{\phi}_{d} = [e^{j \theta_{1}}, \ldots, e^{j \theta_{K}} ]^{T} \in {\mathbb C}^{K}$ is the phase-shift vector of the LIS during the data transmission phase where, $ 0\leq \theta_{k} \leq 2\pi, \; k=1,\ldots,K $ are the phase shifts introduced by the $K$ LIS elements, $\bm{w} \in {\mathbb C}^{M}$ is the beamforming vector at the BS, $x_d \in {\mathbb C}$ is the information bearing signal with ${\mathbb E}[|x_d|^2]=1$, $\sqrt{P_{tx}}$ is the transmit power and $n_d \sim {\mathcal CN} (0,\sigma_d^2)$ is the AWGN at the UE. The phase-shift vector and the beamforming vector are selected based on \cite{mishra2019channel}, which proposed a low complexity design that maximizes the received SNR at the UE,
\begin{equation}
    \gamma_r=\frac{P_{tx}}{\sigma_d^2}|(\bm{h}_{d}^{T} + \bm{\phi}_{d}^{T}\bm{V}^T)\bm{w}|^2 \; .
    \label{ed2}
\end{equation}
Let $\hat{\bm{h}}_d$ and $\hat{\bm{V}}$ be the estimates of the UL channels $\bm{h}_d$ and $\bm{V}$ respectively. Then, using these as plug-in estimates for the optimal $\bm{w}$ and $\bm{\phi}_{d}$ that maximize (\ref{ed2}) \cite{mishra2019channel} leads to the beamformers 
\begin{subequations}
\begin{equation}
    \bm{\phi}_{d_{{\rm opt}}}= {\rm exp}\{-j \phase{\hat{\bm{V}}^{T}\hat{\bm{h}}_d^{*}}\} \; ,
    \label{ed3.1}
\end{equation}
\begin{equation}
    \bm{w}_{{\rm opt}}= \frac{\hat{\bm{h}}_d^{*}+ \hat{\bm{V}}^{*} \bm{\phi}_{d_{{\rm opt}}}^{*}}{ ||\hat{\bm{h}}_d^{*}+ \hat{\bm{V}}^{*} \bm{\phi}_{d_{{\rm opt}}}^{*} ||} \; .
     \label{ed3.2}
\end{equation}
\end{subequations}
With these choices, the achievable rate at the UE is given by 
\begin{equation}
R= \left(1-\frac{T_p}{T_c} \right) \log_2\left(1+ \Bar{\gamma}|(\bm{h}_{d}^{T} + \bm{\phi}_{d_{{\rm opt}}}^{T}\bm{V}^T)\bm{w}_{{\rm opt}}|^2 
 \right)
 \label{ed4}
 \end{equation}
where $\Bar{\gamma}=P_{tx}/\sigma_d^2$ is the average transmit SNR, $T_c$ is the channel coherence time, and we recall that $T_p$ is the pilot duration.
\begin{figure}[htp] 
\centering
\subfigure[$K=50$]{%
\includegraphics[width=0.5\textwidth]{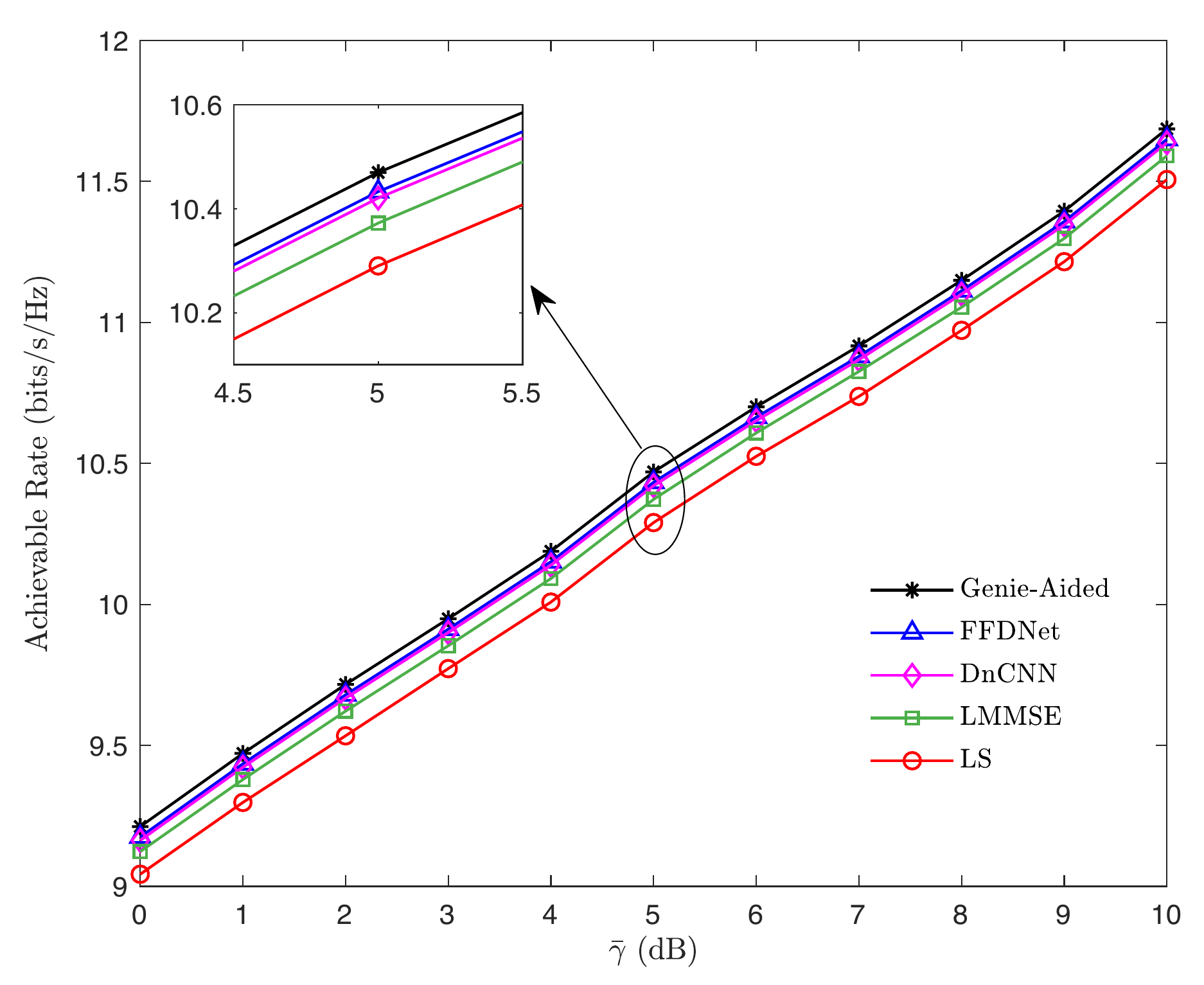}%
\label{fig4.3:a}%
}\hfil
\subfigure[$K=10$]{%
\includegraphics[width=0.5\textwidth]{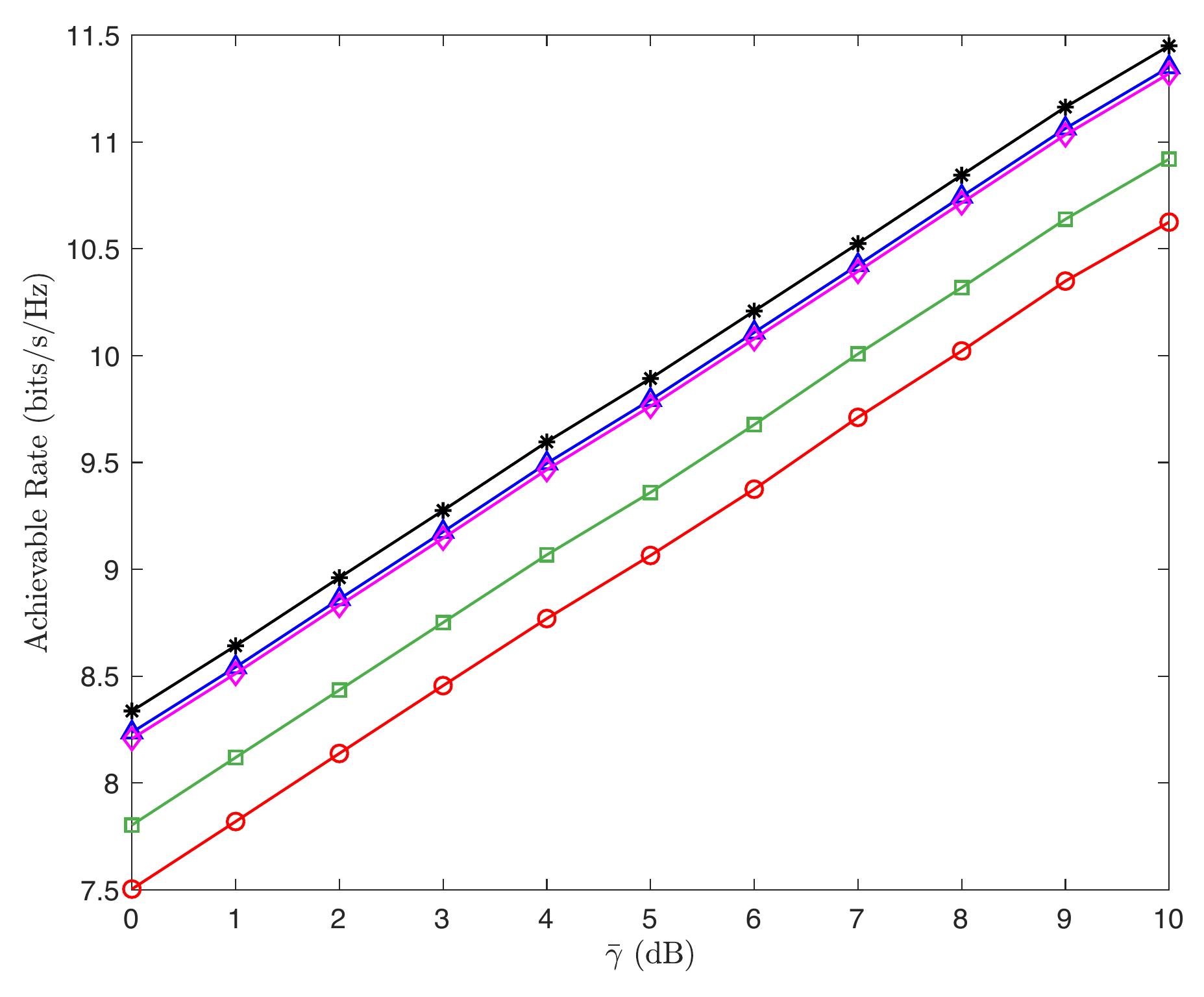}%
\label{fig4.3:b}%
}
\caption{The plots show the achievable rate at the UE versus the average transmit SNR, $\Bar{\gamma}$. Results shown for $M=10$, and two different values of $K=10,50$, using different channel estimation schemes for designing the beamforming vectors. The `Genie-Aided' curve corresponds to the case where the channels are known perfectly. }
\label{fig4.3}
\end{figure}

We study the achievable rate at the UE for the different channel estimation schemes. We adopt the same system parameters as before, though here we assume a training SNR of $\gamma_{\rm tr} = -10$ dB, and consider a channel coherence time $T_c=196$. The achievable rates (\ref{ed4}), with the beamforming vectors (\ref{ed3.1}) and (\ref{ed3.2}) computed based on each of the channel estimation schemes, are shown in Figs. \ref{fig4.3:a} and \ref{fig4.3:b} as a function of the transmit SNR $\bar{\gamma}$. As before, the downlink data transmission performance of LMMSE channel estimator is shown only for the DFT based design. 

We see that the achievable rate is maximum for the beamforming vectors designed using the DnCNN and FFDNet based channel estimators, with their performance approaching to that of the beamforming vectors constructed based on perfect knowledge of the channels (i.e., the `Genie-Aided' channel estimator). The performance improvement of the DnCNN and FFDNet based  estimators over the linear estimators is most evident for smaller numbers of LIS elements, $K$.

\begin{figure}[htp] 
\centering
\includegraphics[width=0.5\textwidth]{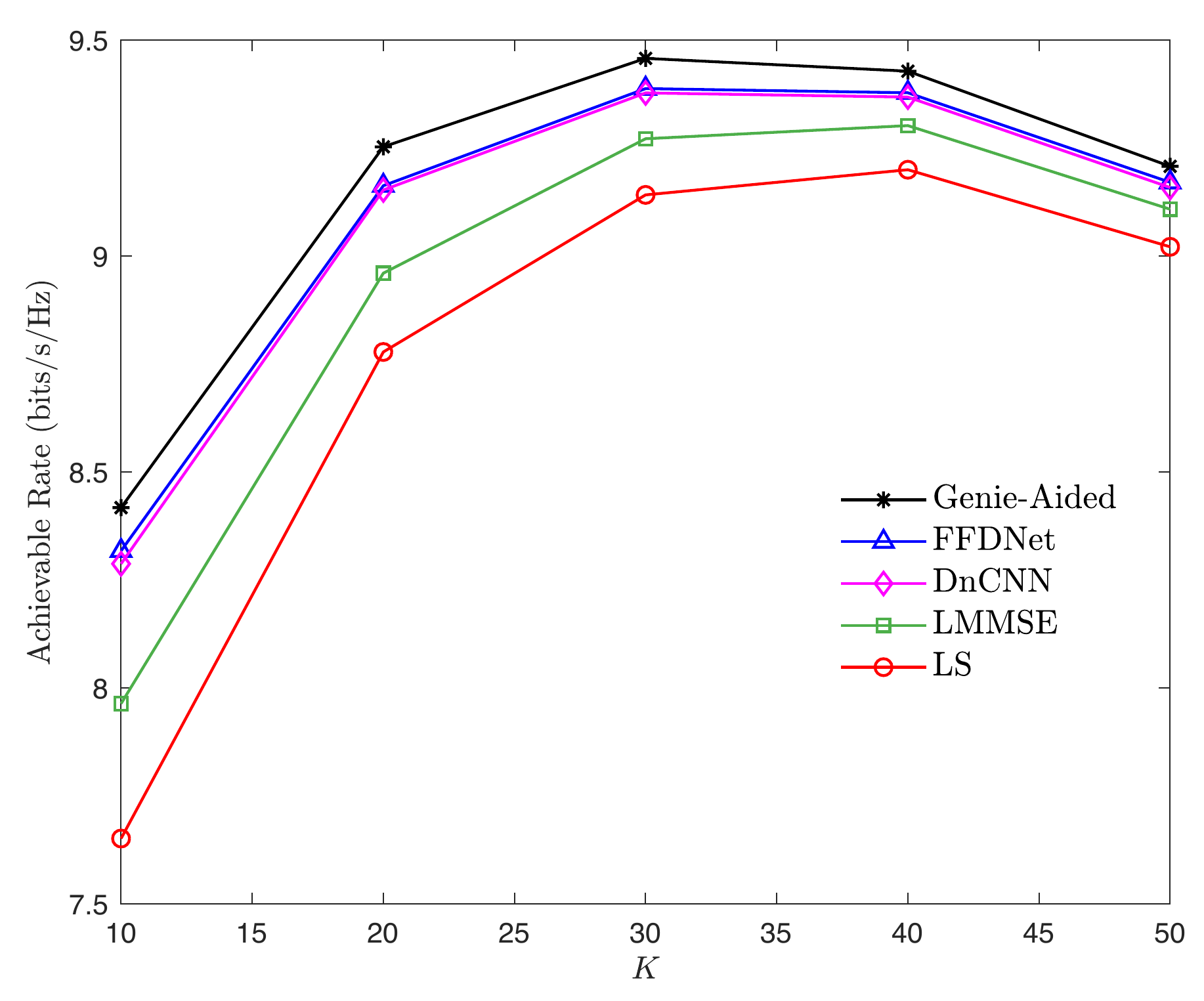}%
\caption{The plots show the achievable rate at the UE versus the number of LIS elements $K$, using different channel estimation schemes for designing the beamforming vectors. Results shown for $M = 10$ and $\Bar{\gamma} = 0\, {\rm dB}$. The `Genie-Aided' curve corresponds to the case where the channels are known perfectly.}
\label{fig4.4}
\end{figure}

Next, we more closely study the effect of increasing $K$ on the achievable rate of the system. As $K$ increases, the received SNR scales with $K^2$ \cite{basar2019wireless,zhou2020spectral}, but at the same time there is a penalty in the pre-log factor of $(\ref{ed4})$ due to the channel estimation overhead, since the pilot duration $T_p=K+1$ increases with $K$. Fig. \ref{fig4.4} shows the achievable rate as the number of LIS elements $K$ changes, with fixed $\Bar{\gamma} = 0\, {\rm dB}$ and $\gamma_{{\rm tr}}= -10\, {\rm dB}$. It is observed that first the achievable rate increases with $K$ due to the increasing beamforming gain from the LIS elements (proportional to $K^2$); however, after reaching a maximum value, the achievable rate starts decreasing. This is because, at large values of $K$, the channel estimation overhead penalty in the pre-log factor of $(\ref{ed4})$ dominates the beamforming gain. The above observation is consistent with a result reported in \cite{kundu2020lis}, where the authors demonstrated a trade-off between achievable rate and the number of LIS elements for a LIS-aided single-antenna system, while similarly accounting for channel estimation overhead. 

\section{Conclusion} \label{conclusion}

In this paper, we have proposed channel estimators for LIS -assisted MISO communications based on the MMSE criterion. First, we derived the best linear estimator, the LMMSE, that admits a closed-form representation depending on the second order statistics of the channels and the noise. The performance of the LMMSE filter depends on the choice of the phase shift matrix employed of the LIS, and hence, we proposed an MM-based algorithm to optimize these phase shifts to yield the smallest MSE. The MM-based algorithm was shown to have good convergence properties, and was used to benchmark the performance of the simple DFT-based phase shifts, which have been shown previously to be optimal for the simple LS-based channel estimators.  We demonstrated that the DFT-based phase shifts constituted a local optimal in the space of feasible phase shift matrices, and produced very similar performance to that achievable with other local optimal attained using the computational MM approach. This, in turn, provides practical motivation for using the DFT-based phase shift matrix with the LMMSE filter. 

Due to the non-Gaussianity of the effective channels in the MISO LIS-aided system, the LMMSE channel estimator is not the globally optimal MMSE solution.  While the globally optimal solution appears analytically intractable, we have proposed two CNN-based image denoising machine learning architectures to approximate this solution. The performance gains of the CNN-based channel estimators over the linear estimators (i.e., LMMSE as well as simple LS) were shown to be quite substantial. The computational cost is also low, particularly since the neural networks can be trained in an offline manner. 

Generally speaking, the problem that we address is well suited to employing a computational neural network solution, since it does not admit a tractable analytical form and the solution is  non-linear. Our numerical comparisons between the neural network based solutions and the LMMSE solution demonstrates that the non-linearity captured by the CNN-based estimators can substantially improve performance, and suggests that such non-linear schemes may be preferable in the context of LIS-aided wireless communication systems. Our results, in general, contribute to the increasing body of work focused on applying deep learning techniques to solve complex optimization problems in wireless communication systems \cite{he2019model,qin2019deep,zappone2019model,chen2019artificial,jin2019channel,he2016deep,dong2019deep,soltani2019deep,gao2019ffdnet}. 

\section{Acknowledgement} \label{ack}
We thank Daniel Palomar for discussions about MM optimization that was helpful for developing the numerical algorithm in Section III-A.








\bibliographystyle{IEEEtran}
\bibliography{IEEEabrv,LIS_Ch}

\end{document}